\documentclass[aps,prc,twocolumn,showpacs,showkeys,preprintnumbers,amsmath,amssymb,a4paper]{revtex4-1}
\usepackage{epsfig}
\usepackage{braket} 
\usepackage{graphicx}

\def\lsim{\mathrel{\rlap{\lower4pt\hbox{\hskip1pt$\sim$}}
    \raise1pt\hbox{$<$}}}                
\def\gsim{\mathrel{\rlap{\lower4pt\hbox{\hskip1pt$\sim$}}
    \raise1pt\hbox{$>$}}}                

\begin{document}
\title{
Faddeev approach to the reaction $K^- d \to \pi \Sigma n$ at $p_{K} = 1$~GeV/c
}

\author{K. Miyagawa$^{1}$, J. Haidenbauer$^{2}$, and H. Kamada$^{3}$  }

\affiliation{
$^{1}$Graduate school of Science, Okayama University of Science, 
1-1 Ridai-cho, Okayama 700-0005, Japan
\\
$^{2}$Institute for Advanced Simulation, Institut f\"ur Kernphysik, 
and J\"ulich Center for Hadron Physics,
Forschungszentrum J\"ulich, D-52425 J\"ulich, Germany
\\
$^{3}$Department of Physics, Faculty of Engineering, Kyushu Institute of Technology, 
KitaKyusyu 804-8550, Japan
}

\begin{abstract}
The reaction $K^-d \to \pi \Sigma n$ is studied within a Faddeev-type approach, 
with emphasis on the specific kinematics of the E31 experiment at 
J-PARC, i.e. $K^-$ beam momentum of $p_K = 1$~GeV/c and neutron angle of 
$\theta_n=0^\circ$. 
The employed Faddeev approach requires as main input amplitudes for the 
two-body subsystems $\bar KN \to \bar KN$ and $\bar KN \to \pi\Sigma$.
For the latter results from recently published chiral unitary models
of the $\bar KN$ interaction are utilized.  
The $\bar KN \to \bar KN$ amplitude itself, however, is taken from a recent 
partial-wave analysis. Due to the large incoming momentum of the $K^-$,
the $\bar KN$ interaction is probed in a kinematical regime  
where those chiral potentials are no longer applicable. 
A comparison of the predicted spectrum for various $\pi \Sigma$ charge
channels with preliminary data is made and reveals a remarkable agreement 
as far as the magnitude and the line shape in general is concerned. 
Noticeable differences observed in the $\pi\Sigma$ spectrum around the $\bar KN$ 
threshold, i.e. in the region of the $\Lambda$(1405) resonance, indicate a 
sensitivity to the details of the employed $\bar KN \to \pi\Sigma$ 
amplitudes and suggest that pertinent high-precision data could indeed 
provide substantial constraints on the structure of the $\Lambda$(1405). 
\end{abstract}


\maketitle

\section{Introduction}
\label{sec:intro}

Modern and systematic approaches that exploit the (approximate) chiral 
and SU(3) flavor symmetries of the underlying QCD Lagrangian have improved 
significantly our understanding of the $\bar K N$ interaction for energies 
in the vicinity of its threshold, see 
Refs.~\cite{Hyodo:2012,Cieply:2016,Kamiya:2016} for recent overviews. 
Nonetheless, some essential questions 
remain. One of them is the detailed pole structure of the $\Lambda$(1405), 
a resonance which is located below but not far from the $\bar K N$ threshold. 
Another, and closely connected topic is the possible existence of (so-called) 
quasi-bound states of the $K^- NN$ system \cite{Sada:2016,Sekihara:2016} and/or 
of kaons with heavier nuclei. A summary of predictions and references 
to the various works can be found in \cite{Gal:2016,Shevchenko:2016}.  

Though chiral SU(3) dynamics provides strong constraints on the $\bar K N$ 
interaction there are still fairly large differences in 
the actual results/predictions, as one can easily see from scanning
through the pertinent literature. This reflects the complexity of the 
underlying physics and is due to the fact that $\bar K N$ cannot be considered
as an isolated system. Possible couplings to the $\pi \Lambda$ and $\pi \Sigma$ 
systems, whose thresholds are just about $100-200$ Mev lower, strongly 
influence the dynamics. 
Most of the available experimental information comes from studies of 
$K^-p$ induced reactions ($K^-p$ elastic scattering, $K^-p\to \bar K^0 n$,
$K^-p \to \pi^0\Lambda$, and $K^-p \to \pi\Sigma$). 
Thus, only isospin combinations of the amplitudes are constrained
by data, but not the individual amplitudes themselves.
As a consequence, there are large variations between the 
isospin $I=1$ $\bar KN$ ($K^- n$) amplitudes around and below the threshold, 
as exemplified, e.g., in Fig.~2 of Ref.~\cite{Cieply:2016}, despite that
all considered interactions are constrained from chiral SU(3) dynamics.  
Actually even for $K^-p$ there is agreement only for energies at and above
the threshold (cf. the same figure), owing to experimental information on 
the level shift and width of kaonic hydrogen \cite{Bazzi:2011}, and the
aforementioned data for $K^-p$ elastic scattering.
The differences in the energy dependence below the threshold reflect variations
in the position of the two poles that are a characteristic feature of the 
$\Lambda$(1405) within chiral approaches 
\cite{Oller:2000,Oset:2002,Ikeda:2012,Guo:2012,Mai:2012,Ohnishi:2013}
(but appear also in conventional meson-exchange dynamics 
\cite{MG,Hai:2011,Kamano-Nakamura-Lee-Sato}). Here specifically 
the pole with the lower mass
is prone to the very details of how chiral SU(3) dynamics is implemented
and has been predicted to be basically anywhere between the $\pi \Sigma$ and
$\bar K N$ thresholds \cite{Hyodo:2012,Cieply:2016,Kamiya:2016}.  

Currently there are major experimental efforts to provide further
constraints on the $\bar K N$ interaction. One of them concerns 
plans for measuring the level shift of $K^- d$ atoms in order to 
pin down the $I=1$ $\bar KN$ amplitude \cite{Zmeskal:2015}.  
Access to the energy dependence of the amplitudes below the $\bar K N$ 
threshold and that means to quantitative information on the pole structure 
of the $\Lambda$(1405) is possible in studies of the $\pi \Sigma$ system. 
Several experiments with that aim have been already performed over the
last few years. Specifically, this concerns measurements of the $\pi\Sigma$ 
invariant mass spectrum in 
photon- \cite{Moriya:2013,Moriya:2013A} and electron induced \cite{Lu:2013} production 
on the proton, in the reaction $pp \to pK^+ \pi \Sigma$ \cite{Zychor:2007,Agakishiev:2012},
and, finally, in $K^-$ induced reactions on a proton \cite{Prakhov:2004} or
deuteron target \cite{JparcE31,JparcE31A}. 

\begin{figure}[t]
\includegraphics[width=7cm]{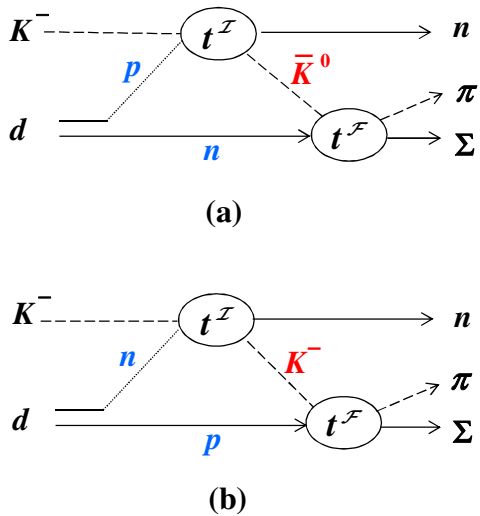}
\caption{Mechanisms considered in the calculation of the reaction
$K^- d \to \pi\Sigma n$: 
(a) $\bar K^0 n \to \pi \Sigma$ rescattering, 
(b) $K^- p \to \pi \Sigma$ rescattering. 
}
\label{f:2step}
\end{figure}

In the present work we focus on the reaction $K^-d \to \pi \Sigma n$ which 
is the objective of the E31 experiment at J-PARC \cite{Jparc}. 
The experiment is performed for specific kinematics, namely for a $K^-$ 
beam momentum of $p_K = 1$~GeV/c and a neutron angle of $\theta_n=0^\circ$.  
Preliminary data for the reaction channels 
$K^- d \to \pi^\pm\Sigma^\mp n$, $K^- d \to \pi^0\Sigma^0 n$,
and $K^- d \to \pi^-\Sigma^0 p$ have been already presented at
conferences \cite{JparcE31A} and in proceedings \cite{JparcE31}
and final results are to be expected soon. Thus, there is a strong
motivation to catch up with this development and to perform 
calculations that are sophisticated enough to facilitate a 
sensible confrontation of theoretical predictions with empirical
information. In the context of the E31 experiment this implies that
the three-body character of the reaction has to be accounted for 
and the formalism best suited for that is the one proposed by
Faddeev. 
Indeed, in the past some calculations for $K^-d \to \pi \Sigma n$ 
have been presented based on a Faddeev-type approach 
\cite{Revai:2014,HYP2015,Ohnishi:2016}
whereas others \cite{KM:2012,Jido:2012,Sekihara:2012,Kamano-Lee}, including 
our own initial study \cite{KM:2012}, took into account only the first terms 
in the multi-scattering expansion, which are depicted in Fig.~\ref{f:2step}.  

Among the studies of the reaction $K^-d \to \pi \Sigma n$
the most instructive one so far is certainly the work 
of Kamano and Lee \cite{Kamano-Lee}. Their calculation, based on 
the diagrams depicted in Fig.~\ref{f:2step}, revealed that the 
$\bar K N\to \bar K N$ amplitude that appears in the first step of 
the two-step process ($t^{\cal I}$ in Fig.~\ref{f:2step}) plays an essential 
role. In particular, this study demonstrated that higher partial waves have to be 
included in this amplitude in the calculation of $K^-d \to \pi \Sigma n$. 
Only then reaction cross section with a magnitude comparable to the experiment 
are obtained. 
Truncating the $\bar K N$ amplitude to the $s$-waves, so far done 
in $K^-d \to \pi \Sigma n$ calculations based on chiral potentials, not least
because in general those models generate $s$ waves only, leads to a gross 
underestimation of the empirical information \cite{Ohnishi:2016}. 

In the present $K^- d$ calculation we take this important aspect into account  
and include higher partial waves in the $\bar K N \to \bar K N$ amplitude. 
Furthermore, we go beyond the two-step approximation of our earlier
study \cite{KM:2012} and treat the $\bar K NN$ three-body scattering process rigorously. 
Since we want to investigate in how far the $\pi \Sigma$ invariant
mass spectrum around the $\bar K N$ threshold, is
sensitive to the details of the $\bar K N$ interaction i.e. to the 
structure of the $\Lambda$(1405), we employ different chiral 
potentials taken from the literature, notably the ones of Ciepl\'y and 
Smejkal~\cite{Cieply:2012} and by Ohnishi et al.~\cite{Ohnishi:2016}. 
Both incorporate the so-called Weinberg-Tomozawa term, i.e. the leading-order 
piece of the effective chiral meson-baryon Lagrangian. 
Furthermore, we consider the chiral interaction proposed by Oset,
Ramos and Bennhold \cite{Oset:2002} which we had used in our initial 
studies of the reaction $K^-d \to \pi \Sigma n$ \cite{KM:2012,HYP2015}. 

The paper is structured in the following way:
Our calculations are performed within a Faddeev-type approach and 
the details of the employed formalism are outlined in 
Sect.~\ref{sec:formalism}. 
Further details can be found in two appendices.
In Sect.~\ref{sec:input} we summarize information about the employed
input for the two-body amplitudes $\bar KN\to \bar KN$ and 
$\bar K N \to \pi\Sigma$. Some characteristic results of these 
two-body amplitudes are likewise provided. 
Our results for the reaction $K^-d \to \pi \Sigma n$ are given and
discussed in Sect.~\ref{sec:results}. The paper ends with a
summary. 


\vspace{\baselineskip}
\section{Formulation of the {\boldmath $K^- d \to \pi \Sigma n$} reaction}
\label{sec:formalism}

\subsection{Faddeev equations}
\label{sec:tec}

We derive the $K^- d \to \pi \Sigma\, n$ amplitude based on the Faddeev 
equations for the $\bar KNN-\pi\Sigma N$ coupled system. 
The nucleon which appears in the $\pi\Sigma N$ system can be either of the two 
being present in the $\bar KNN$ system. 
Clearly the $\bar KNN$ part of the wave function has to be antisymmetrical under exchange of 
the two nucleons, hence  the part describing $\pi\Sigma N$ must reflect it.
This feature can be formulated explicitly by the generalized Pauli principle introduced in Ref.~\cite{generalP}.  
\\ 
\indent
Let us first write down the Faddeev equations for a meson numbered 1 and two baryons numbered 2 and 3:
\begin{alignat}{3}
  \ket{\psi^{(23)}} =    &\:\ket{\phi}  &&+  G_0\  t_{23}  && (\,\ket{\psi^{(12)}}+ \ket{\psi^{(13)}}\,) 
\label {Fa1}
  \\
 \ket{\psi^{(12)}} =    &          &&+  G_0\  t_{12}  && (\,\ket{\psi^{(23)}}+ \ket{\psi^{(13)}}\,) 
 \label{Fa2}
\\
 \ket{\psi^{(13)}} =    &          &&+  G_0\  t_{13}  && (\,\ket{\psi^{(23)}}+ \ket{\psi^{(12)}}\,) \, , 
\label{Fa3}
\end{alignat}
\noindent
where $\phi$ indicates an incoming wave, and $t_{ij}$ are the two-body transition operators embedded
in the three-particle space.
The total wave function $\psi$ is the 
sum of the three components:
$$\ket{\psi}=\ket{\psi^{(23)}}+\ket{\psi^{(12)}}+\ket{\psi^{(13)}}  \ .$$

For the $\bar KNN-\pi\Sigma N$ system we introduce particle labels, in addition to the usual 
space and spin labels, in the form of $ \ket{\, a \alpha \beta}$~\cite{generalP}. 
The particle labels denote $ \{  a \alpha \beta \}=\{ KNN, \pi\Sigma N, \pi N \Sigma\}$, and
the state $\ket{\, a \alpha \beta }$ stands for  
$\ket{a}_1\ket{\alpha}_2\ket{\beta}_3$.   
The completeness relation in that particle space is given by 
\begin{equation}
 \ket{\bar KNN}\bra{\bar KNN} + \ket{\pi\Sigma N}\bra{\pi\Sigma N}+\ket{\pi N\Sigma}\bra{\pi N\Sigma} =1 \ .
\nonumber 
\end{equation}

\noindent
Using the basis above, one can construct a fully symmetric Hamiltonian
with regard to two baryons and a corresponding antisymmetric wave function
(For more details, see Ref.~\cite{generalP}).

Let us impose antisymmetry, $P_{23}\ket{\psi}=-\ket{\psi}$, where the operator $P_{23}$ indicates
the permutation of particles 2 and 3. Then we have the antisymmetric relations
between the Faddeev components:
$$P_{23} \ket{\psi^{(23)}}=-\ket{\psi^{(23)}} ,  \hskip10pt
P_{23} \ket{\psi^{(12)}}=-\ket{\psi^{(13)}}  \ .
$$
\\
\indent
Taking particle representations for the Faddeev equations~(\ref{Fa1}-\ref{Fa3})
with the above antisymmteric relations, one can derive
the following coupled  equations for the five independent components:

\begin{widetext}
\begin{alignat}{3}
  \psi^{(23)}_{\bar KNN} =&\:\phi_{K^- d}&&+  G_0({\scriptstyle \bar KNN})\  t_{NN}(23)  &&(1-P_{23}) \ \psi^{(12)}_{\bar KNN}   
\label{pFa1}
  \\
 \psi^{(12)}_{\bar KNN} =&        &&\quad\:  G_0({\scriptstyle \bar KNN}) \    t_{\bar KN, \bar KN} (12)\ && (\ \psi^{(23)}_{\bar KNN} -P_{23}\ \psi^{(12)}_{\bar KNN}  )
 \nonumber\\
                               &        &&+ G_0({\scriptstyle\bar KNN}) \    t_{\bar KN, \pi\Sigma} (12) && (\ \psi^{(23)}_ {\pi\Sigma N} +\ \psi^{(13)}_{ \pi\Sigma N}  ) 
 \label{pFa2}
  \\
\vbox{\vspace{0.5cm}}
 \psi^{(12)}_{\pi\Sigma N} =&     &&\quad\:  G_0({\scriptstyle \pi\Sigma N})\ \  t_{\pi\Sigma, \bar KN} (12) && (\ \psi^{(23)}_{\bar KNN} -P_{23}\ \psi^{(12)}_{\bar KNN}  )    \nonumber\\
                               &         &&+ G_0({\scriptstyle\pi\Sigma N})\ \  t_{\pi\Sigma, \pi\Sigma} (12)&& (\ \psi^{(23)}_ {\pi\Sigma N} +\ \psi^{(13)}_{ \pi\Sigma N}  )
   \label{pFa3}         \\
\psi^{(23)}_{\pi\Sigma N} =&       &&\quad\:  G_0({\scriptstyle\pi\Sigma N})\ \  t_{\Sigma N}(23) &&  (\ \psi^{(12)}_ {\pi\Sigma N} +\ \psi^{(13)}_{ \pi\Sigma N}  )    
   \label{pFa4}         \\
\psi^{(13)}_{\pi\Sigma N} =&         & &\quad\:  G_0({\scriptstyle\pi\Sigma N})\ \  t_{\pi N}(13)  &&  (\ \psi^{(12)}_ {\pi\Sigma N} +\ \psi^{(23)}_{ \pi\Sigma N}  ) \, ,     
   \label{pFa5} 
\end{alignat}    
\end{widetext}

\noindent
where the components are defined as, for example, 
$\psi^{(23)}_{\bar KNN}=\braket{ \bar KNN | \,\psi^{(23)} }$.

It is a standard procedure~\cite{generalP, Gloeckle96} to extract various partial breakup amplitudes from each individual kernel part of 
the set~(\ref{pFa1})-(\ref{pFa5}). 
Those are introduced into Eqs.~(\ref{Fa1})-(\ref{Fa3}) as 
\begin{eqnarray}
  \ket{\psi^{(23)}} =    &\:\ket{\phi}  &+  G_0\  T^{(23)}  \ket{\phi}
\label{tdef1}
\\
\vspace{15pt}
 \ket{\psi^{(ij)}} =    &    &+  G_0\  T^{(ij)} \ket{\phi} ,   \hskip5pt
 \scalebox{0.85}{$\displaystyle  (ij)=(12),(13) \, . $}  \hskip10pt
\label{tdef2}
\end{eqnarray}

Projecting these equations onto particle states,  one can derive
the following  coupled equations for partial breakup amplitudes which have the same 
structure as the set~(\ref{pFa1})-(\ref{pFa5}):

\begin{widetext}
\begin{alignat}{3}
  T^{(23)}_{\bar KNN} =&       && \quad\: t_{NN}(23)\  G_0({\scriptstyle \bar KNN})   &&(1-P_{23}) \ T^{(12)}_{\bar KNN}   
  \label{tFa1}
  \\
 T^{(12)}_{\bar KNN} =& \,t_{\bar KN, \bar KN}  (12)  \ \phi_d
                                    &&  + t_{\bar KN, \bar KN} (12)\ G_0({\scriptstyle \bar KNN}) \:  && (\, T^{(23)}_{\bar KNN} -P_{23}\ T^{(12)}_{\bar KNN}  )
 \nonumber\\
                           &       &&+  t_{\bar KN, \pi\Sigma} (12)\ G_0({\scriptstyle  \pi\Sigma N}) \    && (\, T^{(23)}_ {\pi\Sigma N} +\ T^{(13)}_{ \pi\Sigma N}  ) 
 \label{tFa2}
  \\
\vbox{\vspace{0.5cm}}
 T^{(12)}_{\pi\Sigma N} =&   \,t_{\pi\Sigma , \bar KN} (12)  \ \phi_d
                                   &&+  t_{\pi\Sigma, \bar KN} (12) \  G_0({\scriptstyle \bar KNN})  && (\, T^{(23)}_{\bar KNN} -P_{23}\ T^{(12)}_{\bar KNN}  )    \nonumber\\
                              &         &&+   t_{\pi\Sigma, \pi\Sigma} (12) \ G_0({\scriptstyle\pi\Sigma N}) && (\, T^{(23)}_ {\pi\Sigma N} +\ T^{(13)}_{ \pi\Sigma N}  )
   \label{tFa3}         \\
T^{(23)}_{\pi\Sigma N} =&       &&\quad\:  t_{\Sigma N}(23) \ G_0({\scriptstyle\pi\Sigma N})  &&  (\, T^{(12)}_ {\pi\Sigma N} +\ T^{(13)}_{ \pi\Sigma N}  )    
   \label{tFa4}         \\
T^{(13)}_{\pi\Sigma N} =&         & &\quad\:   t_{\pi N}(13) \ G_0({\scriptstyle\pi\Sigma N})   &&  (\, T^{(12)}_ {\pi\Sigma N} +\ T^{(23)}_{ \pi\Sigma N}  ) \, , 
   \label{tFa5} 
\end{alignat} 
\end{widetext}

\noindent       
where the partial amplitudes are defined as, for example,
$T^{(23)}_{\bar KNN}=\braket{ \bar KNN | \, T^{(23)} | \phi }$,
and $\phi_d$ is the deuteron wave function. 
The breakup amplitude into the ``physical'' $\pi\Sigma N$ channel is obtained by
\begin{align}
&T(
\scalebox{0.85}{$\displaystyle \: K^- d \to \pi\Sigma N \:  $} 
) 
\nonumber\\
&=\braket{ \pi\Sigma N | \frac{1-P_{23}}{\sqrt2} \,  \bigl(\, T^{(12)}+T^{(23)}+T^{(13)} \bigr) |\,\phi }
\nonumber\\
&= \sqrt2\ \  \Big\{ \ T^{(12)}_{\pi\Sigma N}+ T^{(23)}_{\pi\Sigma N}+ 
T^{(13)}_{\pi\Sigma N} \ \Big\}  \:\: .
\end{align}

\subsection{Technicalities and Relativity} 
\label{sec:tec1}
Here we explain some details concerning the numerical treatment and relativity in solving the Faddeev-type equations~(\ref{tFa1})-(\ref{tFa5}).
Let us illustrate them taking one of the kernel parts of Eq.~(\ref{tFa2}), for example, 
\begin{equation}
 T^{(12)}_{\bar KNN} =  t_{\bar KN, \bar KN} (12)\ G_0({\scriptstyle \bar KNN}) \: (-P_{23}\ T^{(12)}_{\bar KNN}  ) \  .
\label{tFaex}
\end{equation}

\noindent
We work in the 3-body center-of-mass (c.m.) system throughout this paper denoting the momentum 
of particle $i$ by $\vec{q}_{\, i} \scalebox{0.9}{ (i=1,2,3)}$,
and use the partial-wave projected momentum space basis 
\begin{equation}
\ket{k\, q\,\, \alpha }\equiv \ket{k\, q\, ; \:  (ls)j\: (\lambda\, s_3) j_s\: J} \, ,
\end{equation}

\noindent
where $\alpha$ indicates various angular momenta in a $jJ$ coupling:
two-body quantum numbers $ (ls)j$, quantum numbers referred to the third particle $(\lambda\, s_3) j_s$, 
and total angular momentum $J$. 
In the nonrelativistic case $k$ and $q$ correspond to the magnitudes of standard Jacobi momenta,
but now in a relativistic generalization, 
 $\vec k$ and $-\vec k$  are momenta of particle 1 and 2
in the c.m. frame of the two-particle subsystem, 
and $\vec q$ indicates the third particle momentum $\vec q_3$.  
The momentum $\vec k$ is related to the 3-body c.m. momenta $\vec{q}_1$ and $\vec{q}_2$ 
in a compact expression \cite{Fong} by

\begin{eqnarray}
 \vec k=\frac{\varepsilon_2\,  \vec{q}_1-\varepsilon_1\, \vec{q}_2}{\varepsilon_1+\varepsilon_2} \, ,
 \label{Eqfong}
 \end{eqnarray}

\noindent
where $\varepsilon_i=(\omega_i +u_i)/2$,
 $\omega_i\equiv\sqrt{q_i^2+m_i^2} $, $u_i\equiv\sqrt{k^2+m_i^2} $.

\indent
Projecting Eq.~(\ref{tFaex}) onto the basis above, we obtain 

\begin{widetext}
\begin{eqnarray}
\braket{  k\, q\, \alpha\, | \, T^{(12)}_{\bar KNN} } = \sum_{\alpha'}  \int  k'^2 dk'   \braket{k\, \alpha | \, t_{12} (q) |\, k' \alpha' } 
\frac{1 } {W-\omega_3 (q)-\omega_{12}(q,k')+i\epsilon}
\nonumber\\
\times  \sum_{\alpha''}  \int k''^2 dk'' \int q''^2 dq'' \braket{ k'  q\, \alpha' |\,P_{23} | \, k'' q'' \alpha'' }  
\: \left( -\braket{ k'' q'' \alpha'' \,   | \, T^{(12)}_{\bar KNN} } \right) \, , 
\label{T12}
\end{eqnarray}
\end{widetext}

\noindent
where $W$ denotes the total energy, and 
$\omega_3(q)=\sqrt{q^2+m_3^2}$,  
$\omega_{12}(q,k') \equiv \sqrt{q^2+W_{12}^2  } $, 
$W_{12}\equiv\sqrt{k'^2+m_1^2} +\sqrt{k'^2+m_2^2}$.
The permutation matrix element $\braket{ k'  q\, \alpha' |\,P_{23} | \, k'' q'' \alpha'' }$ 
on the right hand side describes a rearrangement  
between different types of basis states $\ket{ \, k' q\, \alpha' }$ and
 $P_{23}\ket{ \, k'' q'' \alpha'' }$, which  is evaluated as

\begin{align}
 &\braket{  k'  q\, \alpha' \, |\, P_{23} | \, k'' q'' \alpha'' } = 
 \nonumber\\
 &\int_{-1}^{1} dx \,  \frac{ \delta (q''-\chi )} {q''^2} \,  \frac{ \delta (k''-\pi )} {k''^2} R_{\alpha'\alpha''}(k' q\, x) \, , 
\label{P23}
\end{align} 

\noindent
where $\chi$ and $\pi$ are shifted momenta given by 
\begin{align}
\chi=&\sqrt{k'^2 +(\rho\,q)^2-2k'\rho q\,  x} \ ,
\label{chi}\\ 
\pi=&\sqrt{(\rho'' k')^2 +(1-\rho''\rho)^2 q^2 - 2 \rho'' k' (1-\rho''\rho) q \, x  }  \  .
\label{pi}
\end{align}

\noindent
The derivation of Eq.~(\ref{P23})  including the expressions of $R_{\alpha'\alpha''}(k' q\, x)$, $\rho$ and $\rho''$ are 
given in Appendix~\ref{permutation}.  Note that in the relativistic case, 
$\rho$ and $\rho''$ are no longer constants but depend on 
$k'$, $q$ and $x$.

In Eq.~(\ref{T12}) together with Eq.~(\ref{P23}), four integrations over $k'$, $k''$, $q''$ and $x$ are left, but
we perform the $k''$ and $x$ integrations analytically using the two $\delta$-functions, which enables us 
\cite{Newform}
to avoid 
the complicated singularity pattern (with moving logarithmic singularities) in standard approaches.
The resulting form is

\begin{widetext}
\begin{eqnarray}
\braket{ k\, q\, \alpha\, | \, T^{(12)}_{\bar KNN} } = \sum_{\alpha'}  \int_0^\infty  k' dk'   \braket{ k\, \alpha | \, t_{12} (q) |\, k'\alpha' } 
\frac{1 } {W-\omega_3 (q)-\omega_{12}(q,k')+i\epsilon}
\nonumber\\
\times  \sum_{\alpha''}   \int_{|k'-\, \rho_{\scalebox{0.6}{$\displaystyle +$}} \, q   |} 
^{k'+\, \rho_{\scalebox{0.55}{$\displaystyle -$} } \, q}
    q'' dq''  \frac{1}{\rho \, q\, f_r}   R_{\alpha'\alpha''}(k' q\, x_0)
\: \left( -\braket{ \pi \, \chi \,  \alpha'' \, |T^{(12)}_{\bar KNN} } \right) \ , 
\label{T12f}
\end{eqnarray}
\end{widetext}
\noindent
where $x_0$ and the factor $f_r$, which is special in the relativistic case, are functions of $k'$, $q$ and $q''$,
while $\rho_-$ and $\rho_+$ are functions of $k'$ and $q$.
Those expressions are given in Appendix \ref{integration:x}.
Note that only a simple pole in the $k'$ variable appears positioned at $k_0$ which
satisfies the relation 
$$\sqrt{(W-\omega_3 (q))^2-q^2}=\sqrt{k_0^2+m_1^2}  +\sqrt{k_0^2+m_2^2} \, , 
$$
giving the zero of the energy denominator. This new prescription of keeping the $k'$ and $q''$ integrations,
which was introduced in Sec. 2.2 of \cite{Newform} and is extended here to the relativistic case, 
greatly reduces the numerical complications.  

\begin{figure*}
\includegraphics[width=7cm,angle=-90]{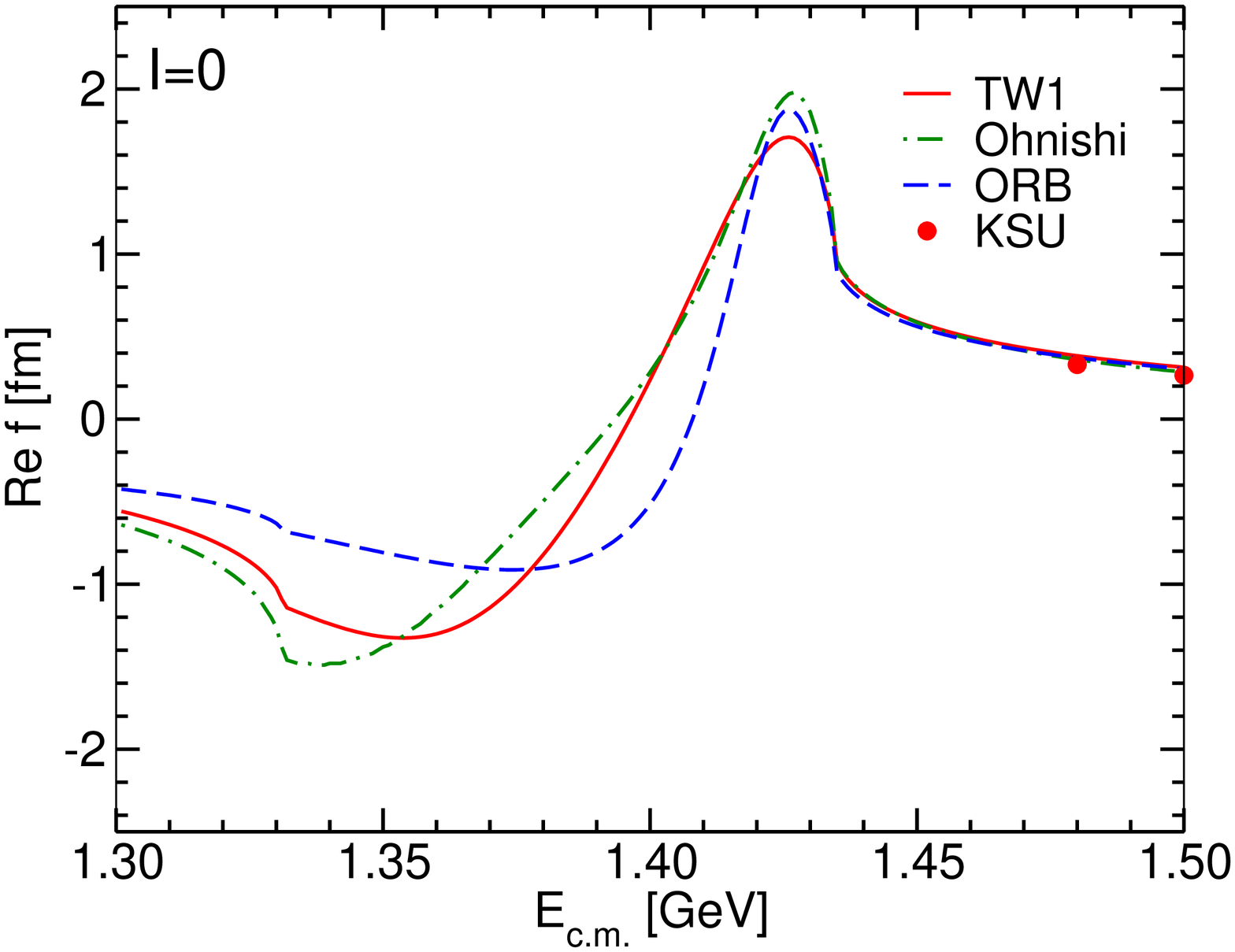}\includegraphics[width=7cm,angle=-90]{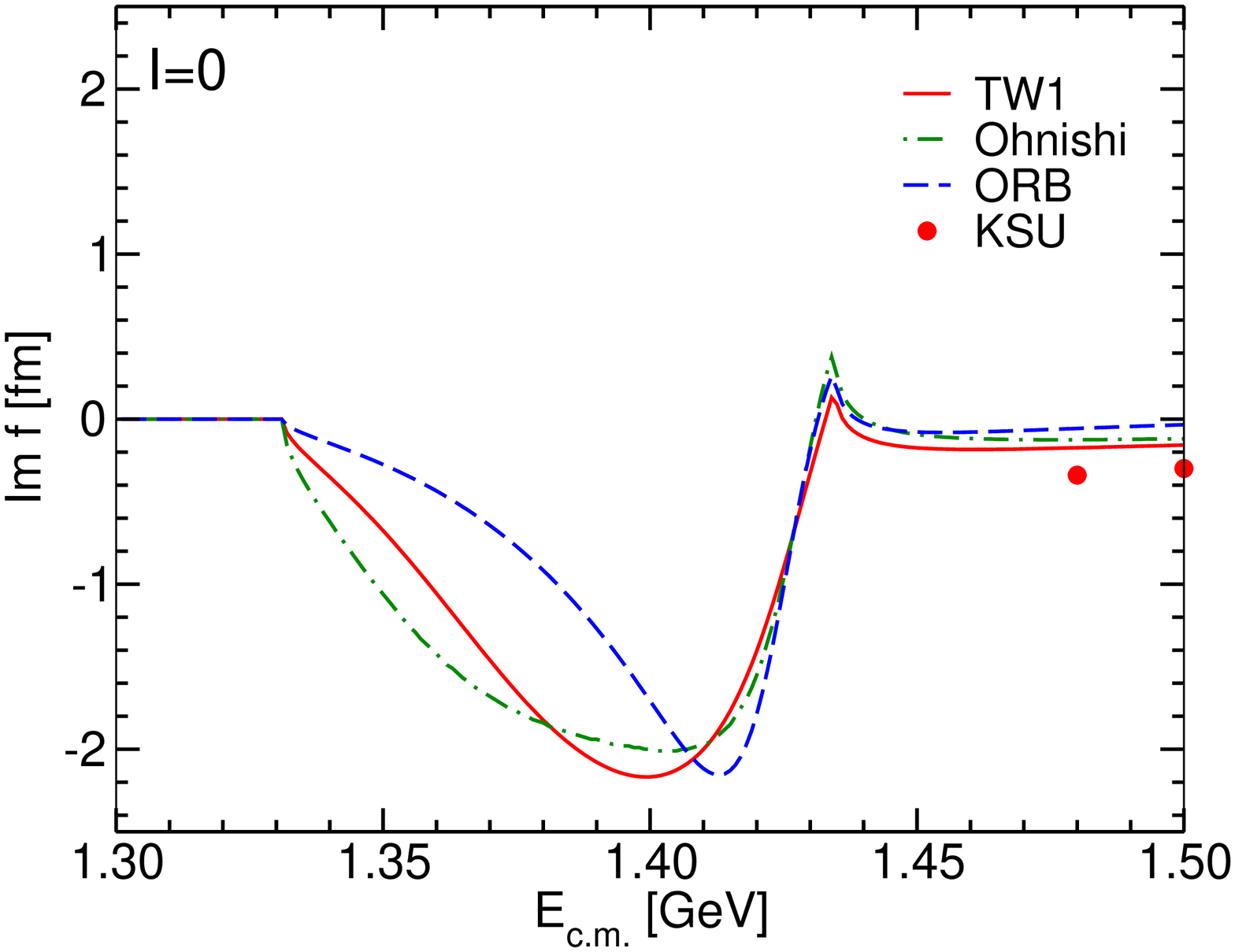}

\includegraphics[width=7cm,angle=-90]{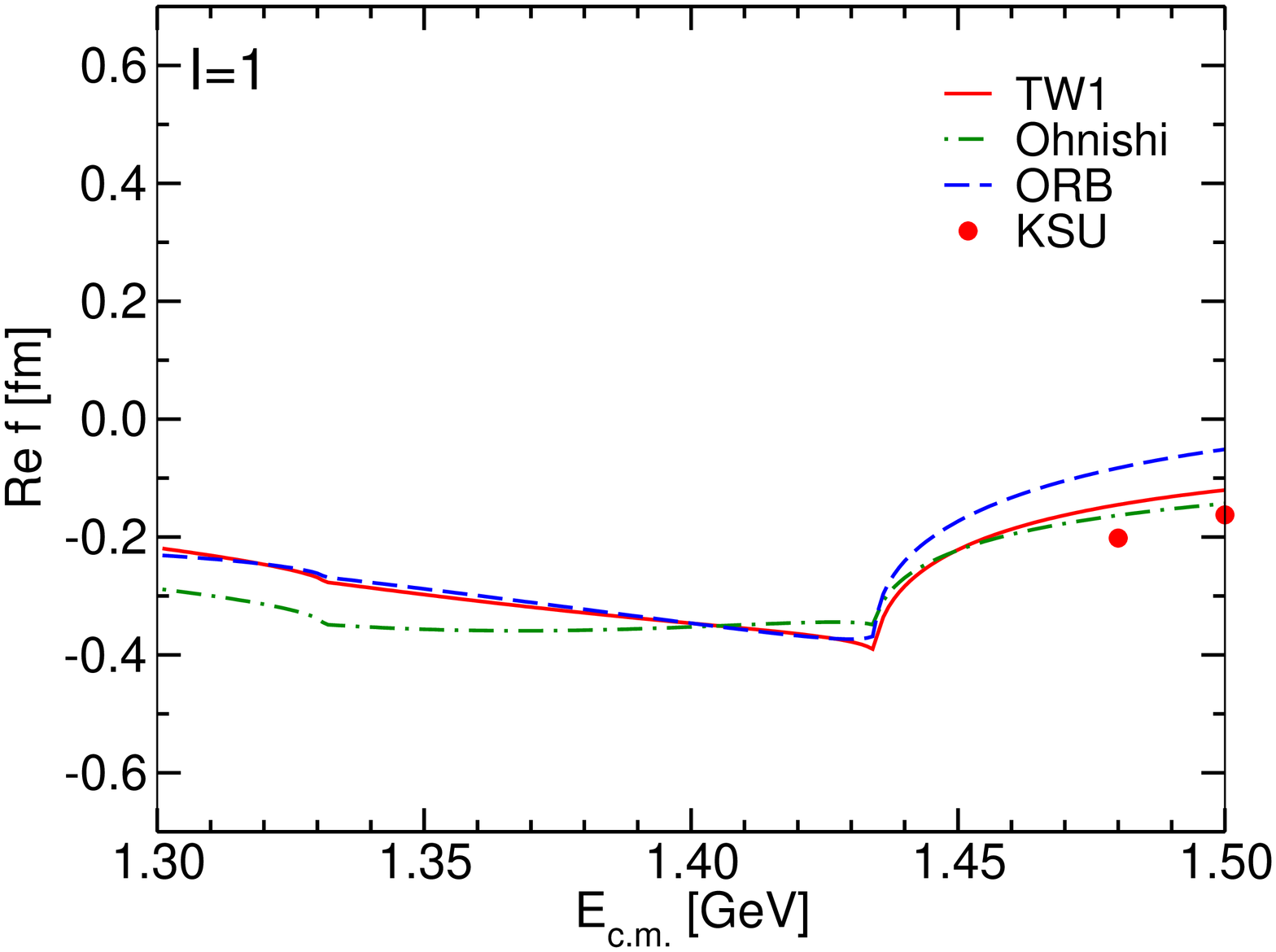}\includegraphics[width=7cm,angle=-90]{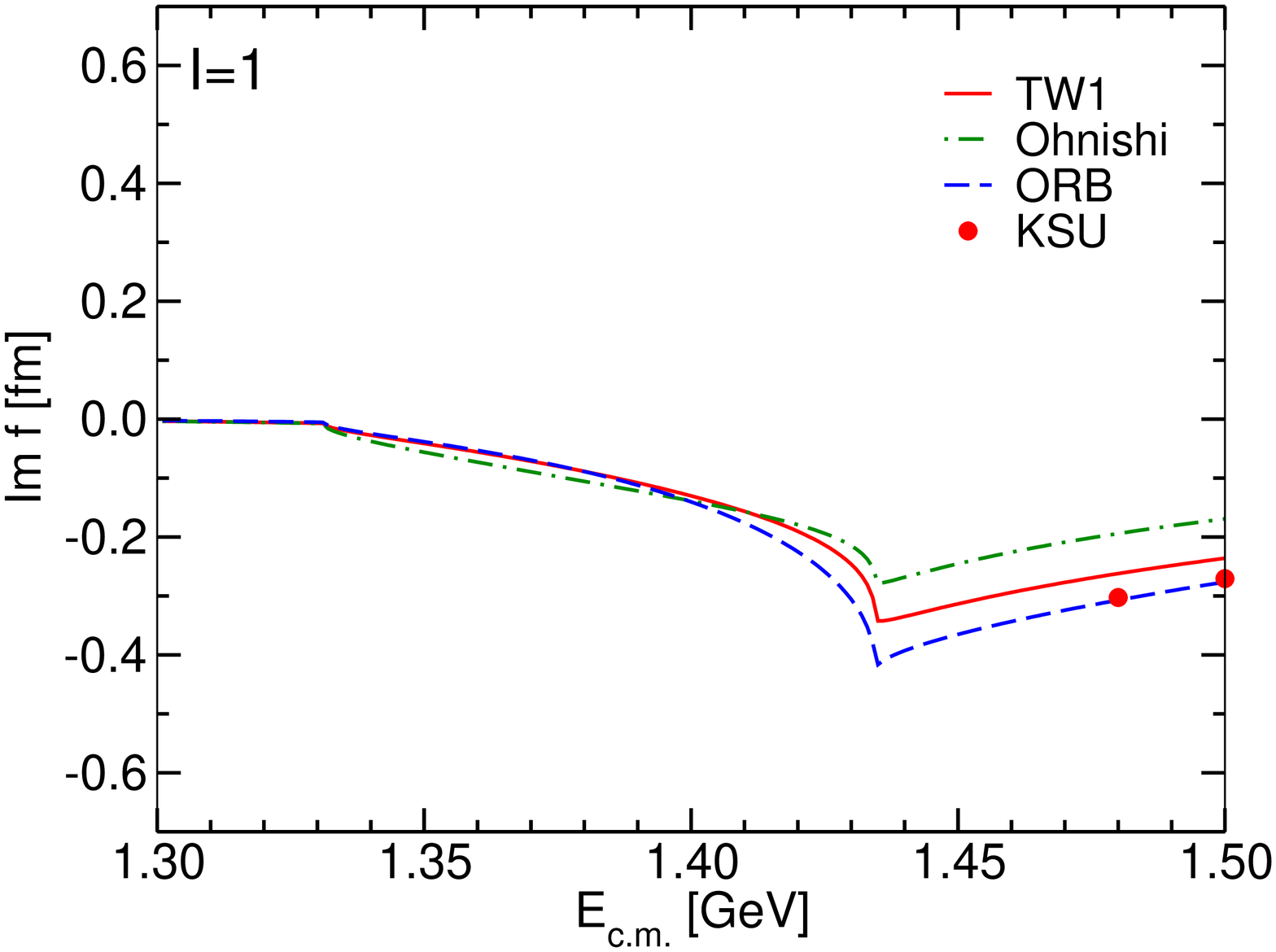}
\caption{$\bar KN\to\pi\Sigma$ isospin $I=0,1$ $s$-wave amplitudes of the employed potentials: 
TW1 \cite{Cieply:2012} (solid lines), Ohnishi et al.~\cite{Ohnishi:2016} (dash-dotted lines), 
and ORB \cite{Oset:2002} (dashed lines), and of the KSU analysis \cite{KSU} (filled circles).
}
\label{f:CMP}
\end{figure*}
 
Finally, we mention that the boosted two-body {$t$-matrix}
$ \braket{  k\, \alpha | \, t_{12} (q )|\, k' \alpha }$
describes transitions in the moving frame with the magnitude of the momentum given by the modulus 
$| \vec{q}_1+\vec{q}_2 |= |\! - \! \vec{q\, } | $.
Following the procedure of a Poincar\'e invariant few-body model developed in \cite{Polyzou}, it is related 
to the $t$-matrix
$ \braket{  k\, \alpha | \, t^{cm}_{12} |\, k' \alpha }$
defined in the c.m. frame of the two particles 1 and 2 as
\begin{eqnarray}
 \braket{  k\, \alpha | \, t_{12} (q )|\, k' \alpha }=\frac{ W_{12} ( k)+W_{12}(k')  } {\omega_{12} (q, k)+\omega_{12} (q, k')}
\braket{  k\, \alpha | \, t^{cm}_{12} |\, k' \alpha } \, , 
\nonumber\\
\label{boost}
\end{eqnarray}
where 
$W_{12}(k)\equiv\sqrt{k^2+m_1^2} +\sqrt{k^2+m_2^2}$, and the
two-body energy for
$\braket{  k\, \alpha | \, t^{cm}_{12} |\, k' \alpha }$
is determined as
$\sqrt{(W-\omega_3 (q))^2-q^2}$.
The expression (\ref{boost}) only holds for half-off-shell $t$-matrices
(see Eq.~(48) of \cite{Polyzou})
but at this stage we use it also for fully off-shell $t$-matrices as an approximation.
The prescription for fully off-shell $t$-matrices has been studied in \cite{Polyzou} and
\cite{Kamada}.


\section{Employed two-body amplitudes} 
\label{sec:input}

There is an abundance of studies of the $\bar KN$ interaction 
that start out from an effective chiral Lagrangian --
either at leading-order or up to next-to-leading order, 
and considering the coupling to the $\pi \Lambda$ and $\pi \Sigma$
channels or even to all meson-baryon systems with strangeness $S=-1$ 
that can be build from the lowest SU(3) (pseudoscalar meson, baryon) 
octets \cite{Hyodo:2012,Cieply:2016,Kamiya:2016,
Oller:2000,Oset:2002,Ikeda:2012,Guo:2012,Mai:2012,Ohnishi:2013}. 
However, only some of the resulting interactions can be readily adapted 
to match with the Faddeev-type three-body approach described above. 
Calculations in that scheme require as input two-body amplitudes 
that are generated from a potential inserted into a standard 
(relativistc or non-relativistic) three-dimensional Lippmann-Schwinger 
equation so that the pertinent reaction amplitudes can be calculated 
for momenta that are on- or off the energy shell. It is worth mentioning that 
the very first study of the $\bar KN$ interaction based on a chiral Lagrangian 
\cite{Kaiser:1995} yielded indeed such a potential. 
 
In the present study we utilize two fairly recent potentials, namely the model 
TW1 (also known as ${\rm P_{WT}}$) by Ciepl\'y and Smejkal \cite{Cieply:2012},  
and the energy-dependent model $V^{E-dep.}$ of Ohnishi et al.~\cite{Ohnishi:2016}.
Both are so-called chirally motivated potentials, i.e. they are based on the 
Weinberg-Tomozawa term, and both yield results in agreement with the latest 
experimental value for the level shift and width of kaonic hydrogen by the 
Siddharta Collaboration \cite{Bazzi:2011}. 
The latter aspect is very important because those data put very
tight constraints on the $K^-p$ scattering length, i.e. on the 
$\bar KN$ interaction close to the threshold. 
The actual expressions for those potentials can be found in 
Refs.~\cite{Cieply:2012} and \cite{Ohnishi:2016}, respectively, 
together with pertinent results for $\bar KN$ and $\bar KN\to \pi\Sigma$ 
(see also Refs.~\cite{Cieply:2010} and \cite{Ohnishi:2013}). 
The formal difference between the two potentials is very small, consisting
only in the treatment of the factors coming from the energies of the
mesons and baryons, cf. Eqs.~(1) \cite{Cieply:2012} and 
(17) \cite{Ohnishi:2016}, respectively. However, the actual fits to the 
data are different and so are the underlying two-body amplitudes. 
Exemplary we show the ones for $\bar KN \to \pi\Sigma$ in Fig.~\ref{f:CMP}. 
Some key results like the ${\bar K N}$ scattering lengths 
and the pole positions of the $\Lambda$(1405) are summarized
in Table~\ref{POLES}.
Besides those interactions we consider also the Oset-Ramos-Bennhold (ORB)
potential~\cite{Oset:2002}. This is done mainly for 
historical reasons. We used this potential in our initial study
of $K^-d \to \pi\Sigma n$ \cite{KM:2012} and we wanted to connect
with those results. Note that the $K^-p$ scattering length predicted
by the ORB potential is not in line with the kaonic hydrogen
results \cite{Bazzi:2011}. 
Still it will be interesting to see in how far the $K^-d\to \pi\Sigma n$
results differ from those for the other potentials. 

\begin{figure}
\includegraphics[width=7cm,angle=-90]{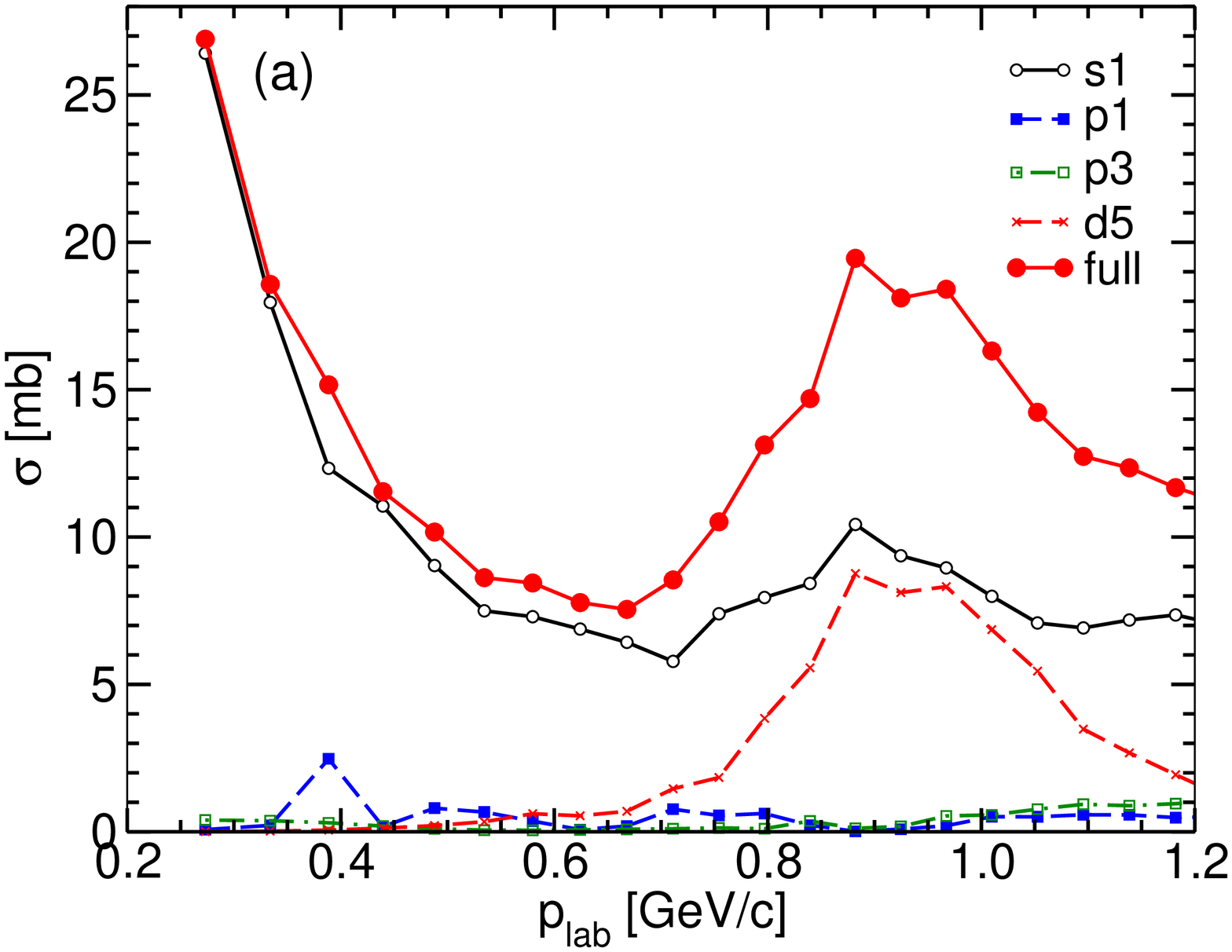}

\includegraphics[width=7cm,angle=-90]{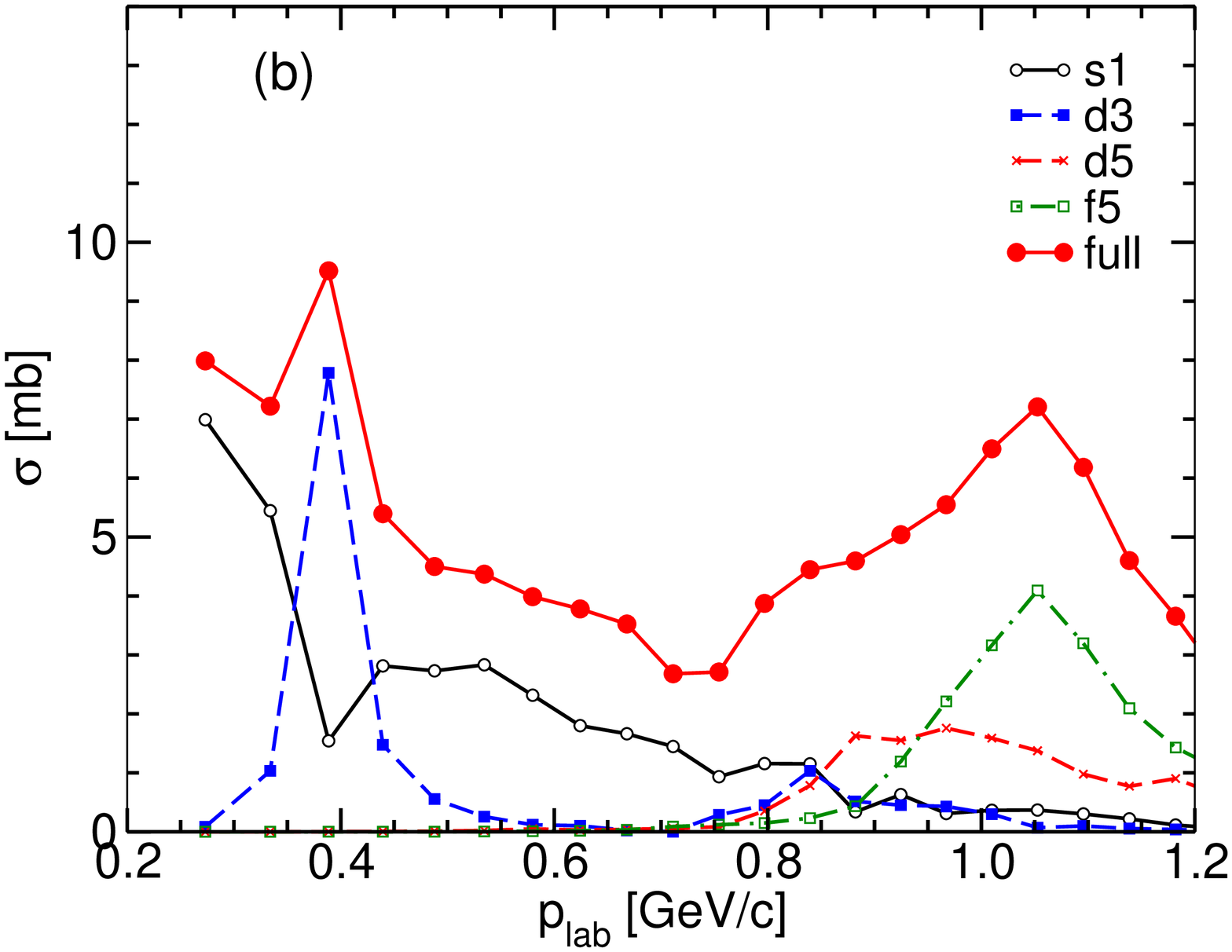}
\caption{Partial-wave cross sections of the KSU analysis \cite{KSU} 
for (a) $K^-n \to K^-n$ and (b) $K^-p \to \bar K^0n$. 
The partial waves are indicated by $L_{2J}$. 
}
\label{f:KSU}
\end{figure}

As already mentioned, typically chiral potentials are limited to $s$-waves only. 
Accordingly, in view of the findings of Kamano and Lee \cite{Kamano-Lee}, 
these are not suitable for generating the $\bar KN$ amplitude that enters into 
the initial scattering process ($t^{\cal I}$ in Fig.~\ref{f:2step}). Thus, in order
to circumvent this difficulty we decided to resort to a phenomenological 
treatment which means that we substitute this amplitude directly by results of a 
partial-wave analysis of available $\bar KN$ scattering data, 
namely the one performed recently by Manley and his group at 
the Kent State University (KSU)~\cite{KSU}. 
This analysis covers the energy range from $1480$ to $2100$ MeV, i.e. goes well 
beyond $p_{K^-}= 1$ GeV/c (corresponding to a $\bar K N$ c.m. 
energy of $1795$ MeV) that is needed for analyzing the E31 experiment.
Taking over those results has the advantages that one implements an amplitude
that yields an excellent reproduction of the $\bar KN$ data and that one can 
use as many partial waves as are needed in the three-body calculation for 
getting converged results for the considered observables. 
Of course, there is a price we have to pay. We have to introduce a 
phenomenological form factor for the required off-shell extension.  
To be concrete, we use the form factor $f(p,k)=\Lambda^2/(\Lambda^2 +(p-k)^2)$, 
which depends on the off-shell momentum $p$ but also on the on-shell momentum $k$. 
This choice ensures that the $\bar KN$ on-shell amplitude remains unchanged. For 
the cutoff mass in the form factor we employ $\Lambda=800$ MeV. However, we performed
test calculations where we varied the value of $\Lambda$ by 10~\% 
in order to examine the sensitivity of our results to this phenomenological 
treatment. Fortunately, it turned out that the variations in the $K^-d$ 
observables that we consider coming from the choice of the cutoff are negligibly small. 

Partial-wave cross sections for $K^-n \to K^-n$ and $K^-p \to \bar K^0n$ of the 
KSU analysis \cite{KSU} are displayed in Fig.~\ref{f:KSU}. Obviously, for 
$p_{K^-} = 1$~GeV/c, i.e. the kinematics of the E31 experiment, there are large 
contributions from the $d_{I\,5}$ and $f_{I\,5}$ amplitudes, respectively. 
(We use here the standard notation $L_{I\,2J}$, but with small letters for the 
two-body amplitudes as it is commonly done in three-body calculations.) 

\begin{table}[h]
\caption{Characteristic results for the considered $\bar KN$ potentials
by Ciepl\'y and Smejkal (TW1) \cite{Cieply:2012}, Ohnishi et al.
($V^{E-dep.}$) \cite{Ohnishi:2016}, and Oset-Ramos-Bennhold (ORB)
\cite{Oset:2002}.
Scattering lengths are in fm and pole positions in MeV.
}
\label{POLES} 
\vskip 0.2cm
\renewcommand{\arraystretch}{1.2}
\centering
\begin{tabular}{|l|c|c|c|c|}
\hline
model  &  $a_{\bar K N}$ ($I=0$) & $a_{\bar K N}$ ($I=1$)  &      pole 1   &   pole 2 \\
\hline
TW1        &  -1.61 +{\rm i}  1.02 & 0.60 +{\rm i}  0.50 & 1433 -{\rm i} 25 & 1371 -{\rm i} 54 \\
$V^{E-dep.}$ &  -1.89 +{\rm i}  1.11 & 0.45 +{\rm i}  0.53 & 1429 -{\rm i} 15 & 1344 -{\rm i} 49 \\
ORB        &  -1.72 +{\rm i}  0.89 & 0.52 +{\rm i}  0.64 & 1426 -{\rm i} 16 & 1390 -{\rm i} 66 \\
\hline
\end{tabular}
\end{table}

For the deuteron wave function, $\phi_d$, we use the one of the Nijmegen potential
Nijm93~\cite{Nij93NN}. We tested wave functions from other realistic potentials
too, but it turned out that the results are rather insensitive to the particular choice. 


\vspace{\baselineskip}
\section{Results and Discussion}
\label{sec:results}

Before presenting the results, we briefly review our earlier studies on the 
$K^-d\to \pi\Sigma N$ reaction and relevant works.
We started with a calculation of the two-step processes~\cite{KM:2012}
for the beam momentum $p_{K}=0.6$ GeV/c and the neutron angle $\theta_n=0^\circ$,
where the $s$-waves of the J\"ulich meson-exchange model \cite{MG}
and the ORB chiral interaction \cite{Oset:2002} were used for the $\bar KN-\pi\Sigma$ 
amplitude. The diagrams included in these calculations are depicted in Fig.~\ref{f:2step}. 
(The plane-wave impulse process, see Fig.~3(A) in Ref.~\cite{KM:2012}, gives negligible 
effects and is not shown.) 
However, no clear peak was seen in the  $\Lambda(1405)$ resonance region, and then
we proceeded to a Faddeev calculation which enabled us to sum up all rescattering
processes. We performed calculations~\cite{HYP2015}
for $p_{K^-}=1$~GeV/c and $\theta_n=0^\circ$ considering the kinematics of the J-PARC
E31 experiment~\cite{Jparc}, and obtained converged results after the sixth 
iteration of Eqs.~(\ref{tFa1})-(\ref{tFa3}), 
where the same $s$-wave $\bar KN-\pi\Sigma$ interaction mentioned above was used.
(The transition to the $\pi\Sigma N$ system was treated perturbatively.)
Although  a peak corresponding to the $\Lambda(1405)$ resonance appeared there, 
the line shape of the $\pi\Sigma$ invariant mass spectrum did not   
match the preliminary E31 results \cite{JparcE31,JparcE31A,Inoue:2017} and, 
in addition and more disturbingly, its magnitude was five times smaller than 
the experiment.  

Recently, Kamano and Lee~\cite{Kamano-Lee} investigated the reaction $K^-d\to\pi\Sigma n$
as well and they realized the importance of the $\bar KN\to \bar KN$ amplitude that enters 
into the first re-scattering process ($t^{\cal I}$ in Fig.~\ref{f:2step}). 
This amplitude is well constrained because there is a wealth of data in the
high-energy region corresponding to the incoming $K^-$ momentum of $1$ GeV/c. 
Their calculation is based on the two-step processes depicted in Fig.~\ref{f:2step}
and they use amplitudes from their coupled-channel $\bar KN$ 
potential~\cite{Kamano-Nakamura-Lee-Sato} which was developed in the course of a 
comprehensive analysis of $\bar K N$ data up to an energy of $2100$~MeV.
In their calculation, $K^-d$ cross sections of comparable magnitude to the E31 experiment 
were obtained \cite{Kamano-Lee}. 

The results in Ref.~\cite{Kamano-Lee} indicate very clearly that it is crucial 
to use the full amplitude for the initial $\bar KN\to \bar KN$ process, and not only 
the $s$-wave contribution as we \cite{KM:2012,HYP2015} and others \cite{Ohnishi:2016}
did in the past.  
Thus, as already mentioned in Sect.~\ref{sec:input}, in the present work we adopt 
the $\bar KN\to \bar KN$  amplitude established by the KSU group~\cite{KSU} which 
describes the $\bar KN$ reaction data in the high-energy region with similar or 
possibly even better quality than the one used in~Ref.~\cite{Kamano-Lee}. 
 
\begin{figure}[t]
\includegraphics[width=7cm,angle=-90]{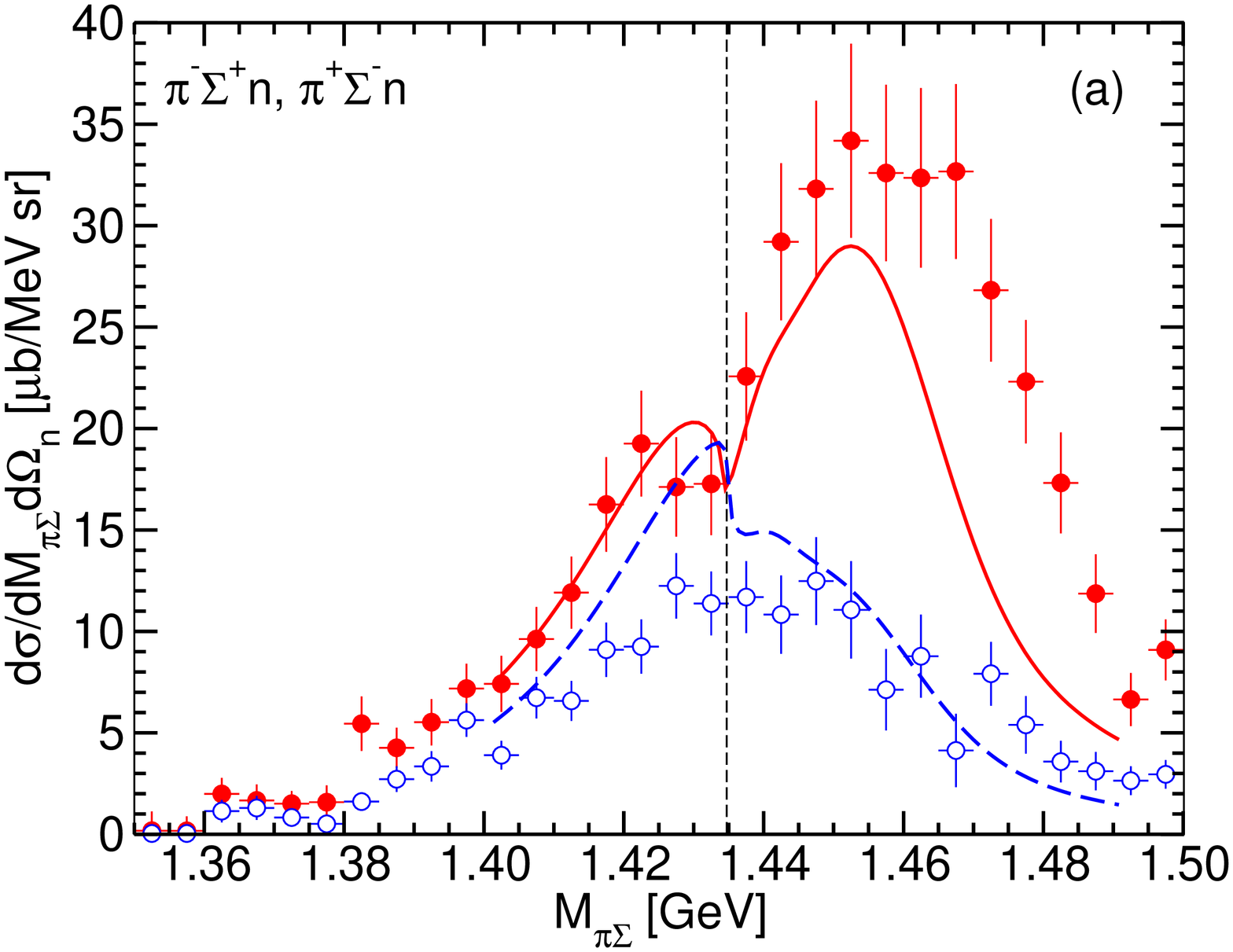}
\includegraphics[width=7cm,angle=-90]{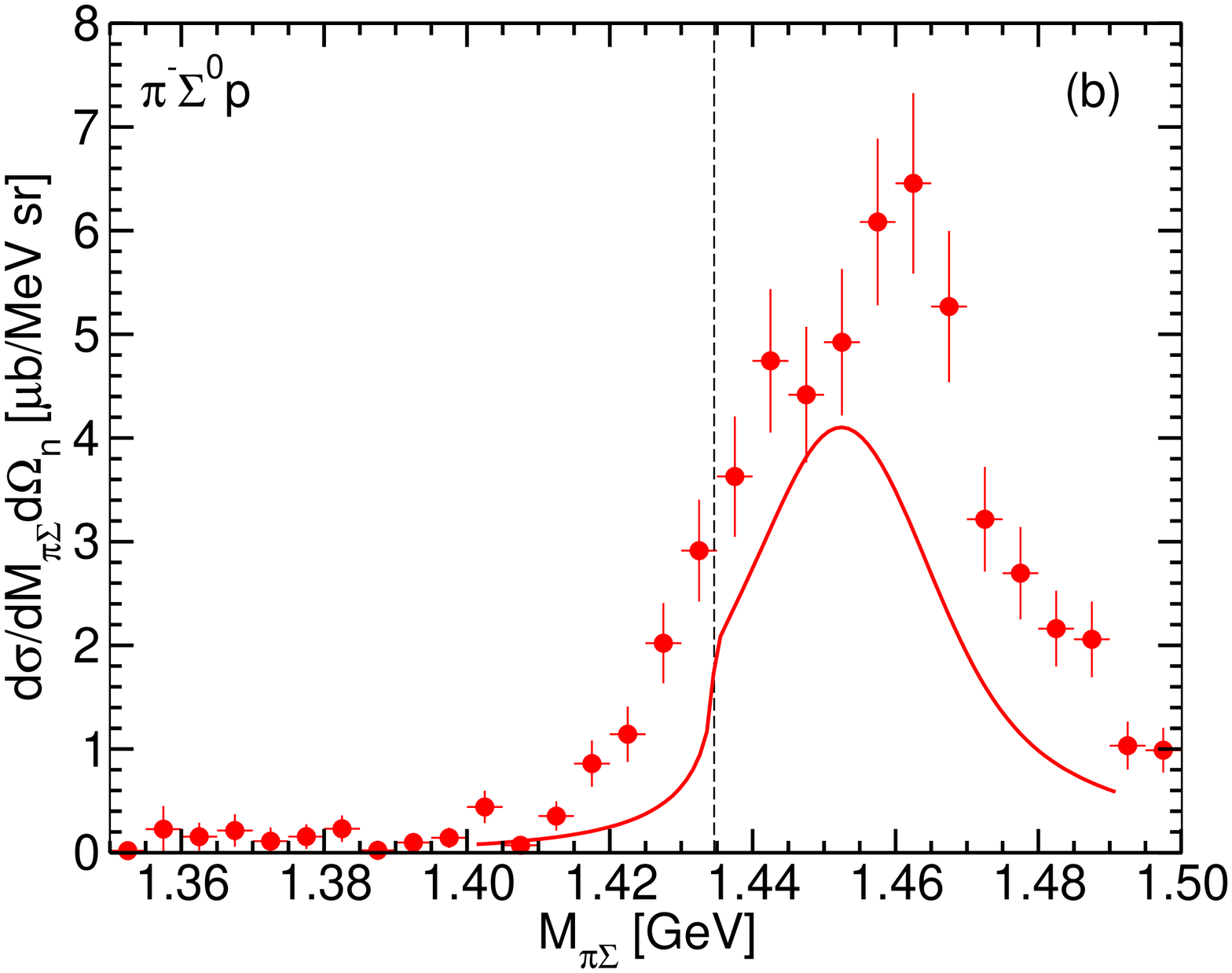}
\caption{Double differential cross section for (a) 
$\pi^\pm\Sigma^\mp n$ and (b) $\pi^-\Sigma^0 p$. Predictions based on the
potential TW1 \cite{Cieply:2012} are indicated by solid 
($\pi^-\Sigma^+ n$, $\pi^-\Sigma^0 p$)
and dashed lines ($\pi^+\Sigma^- n$), respectively. 
Preliminary data are taken from Ref.~\cite{JparcE31A}. 
Vertical line indicate the $\bar K N$ threshold.
}
\label{f:XS1}
\end{figure}

The merits of the special kinematics of the E31 experiment have been discussed 
thoroughly in Refs.~\cite{Jparc,Kamano-Lee}. In short, the $K^-$ kicks out the neutron 
(or proton) from the deuteron and is thereby strongly slowed down. The slowly moving $K^-$ 
interacts then with the remaining nucleon and converts into $\pi\Sigma$. Viewed in the c.m. 
frame the outgoing nucleon and the $\pi\Sigma$ system move back to back \cite{Kamano-Lee}
and, therefore, there is basically no correlation between them.  
Another important aspect is that the energy regions in which the involved two sub-processes,
$\bar KN \to \bar KN$ and $\bar K N \to \pi \Sigma$, take place are well separated for this
special kinematics. Specifically, for $\pi\Sigma$ invariant masses below $1490$ MeV, say, 
i.e. the region of interest where the majority of the E31 data are,
the corresponding energy for $t^{\cal I}$, i.e. the $\bar KN\to \bar KN$ amplitude, 
is essentially above $1550$ MeV or so, and thus well above the $\bar KN$ threshold.
Therefore, there is no principle conflict when using the $\bar KN \to \bar KN$
amplitude from the partial-wave analysis and the $\bar K N \to \pi \Sigma$ amplitude
from chirally motivated potentials. 
Finally, according to Kamano and Lee, only the $s$-wave of the $\bar KN\to \pi\Sigma$ amplitude 
is of relevance for the considered observables, cf. Fig. 7 in Ref.~\cite{Kamano-Lee}. Thus, it 
is meaningful to combine the full $\bar KN \to \bar KN$ amplitude from the KSU analysis and 
the $s$-wave $\bar KN\to \pi\Sigma$ amplitude from chiral potentials. Indeed, variations in
the results reflect directly differences in the $\bar KN\to \pi\Sigma$ amplitude around
and below the $\bar KN$ threshold as predicted by the chiral interactions. 
On the other hand, existing differences in the ($s$-wave) $\bar KN \to \bar KN$ amplitude 
in the near-threshold region \cite{Cieply:2016}, do not play a role for the actual results. 

\subsection{Comparison with preliminary E31 results}
 In Fig.~\ref{f:XS1} inclusive cross sections for the reaction $K^-d\to \pi\Sigma N$ are 
 shown as a function of the $\pi\Sigma$ invariant mass. The $K^-$ beam momentum and 
 the neutron angle are fixed to $p_{K^-}= 1$ GeV and $\theta_n=0^\circ$, respectively,
 in accordance with the J-PARC E31 experiment~\cite{JparcE31}. We use the amplitudes
 by the KSU group~\cite{KSU} for the $\bar KN \to \bar KN$ processes depicted as $t^{\cal I}$ 
 in Fig.~\ref{f:2step}, while the chirally motivated interaction TW1~\cite{Cieply:2012} 
 by Ciepl\'y and Smejkal is utilized for generating the $\bar KN \to \pi\Sigma$ amplitude that 
 are represented by $t^{\cal F}$ in Fig.~\ref{f:2step}. Partial waves up to a total angular 
 momentum of $j=7/2$ are included for the $\bar KN\to\bar KN$ amplitude. 
 Isospin-averaged masses are used so that the $\bar KN$ threshold is at $1434.6$~MeV.
 
 The predicted line shapes for the three final states in Fig.~\ref{f:XS1}
 are compared with available, but still preliminary data of the E31 experiment~\cite{JparcE31A}. 
 In contrast to our former work~\cite{HYP2015} where only an $s$-wave interaction 
 was used for the amplitude of the first-step ($t^{\cal I}$) the cross sections increase
 drastically and reach a magnitude that is comparable to the experiment. 
 The importance of using the full amplitude for $t^{\cal I}$ becomes immediately clear when
 one inspects the $\bar KN\to \bar KN$ cross section generating by the KSU amplitudes, displayed 
 in Fig.~\ref{f:KSU}. It can be seen that the partial waves $d_{I\,5}$ and $f_{I\,5}$ 
 provide large contributions to the cross section around $p_{K} \approx 1$~GeV/c,
 the relevant energy region for the amplitude in the first step. 

Indeed, in our new calculation there is a good overall agreement with the
preliminary data for $K^-d\to \pi^-\Sigma^+ n$. In particular, the maximum of the 
spectrum is roughly reproduced. Qualitatively, the results are similar to those reported by 
Kamano and Lee in Ref.~\cite{Kamano-Lee}. 

\begin{figure*}
\includegraphics[width=7cm,angle=-90]{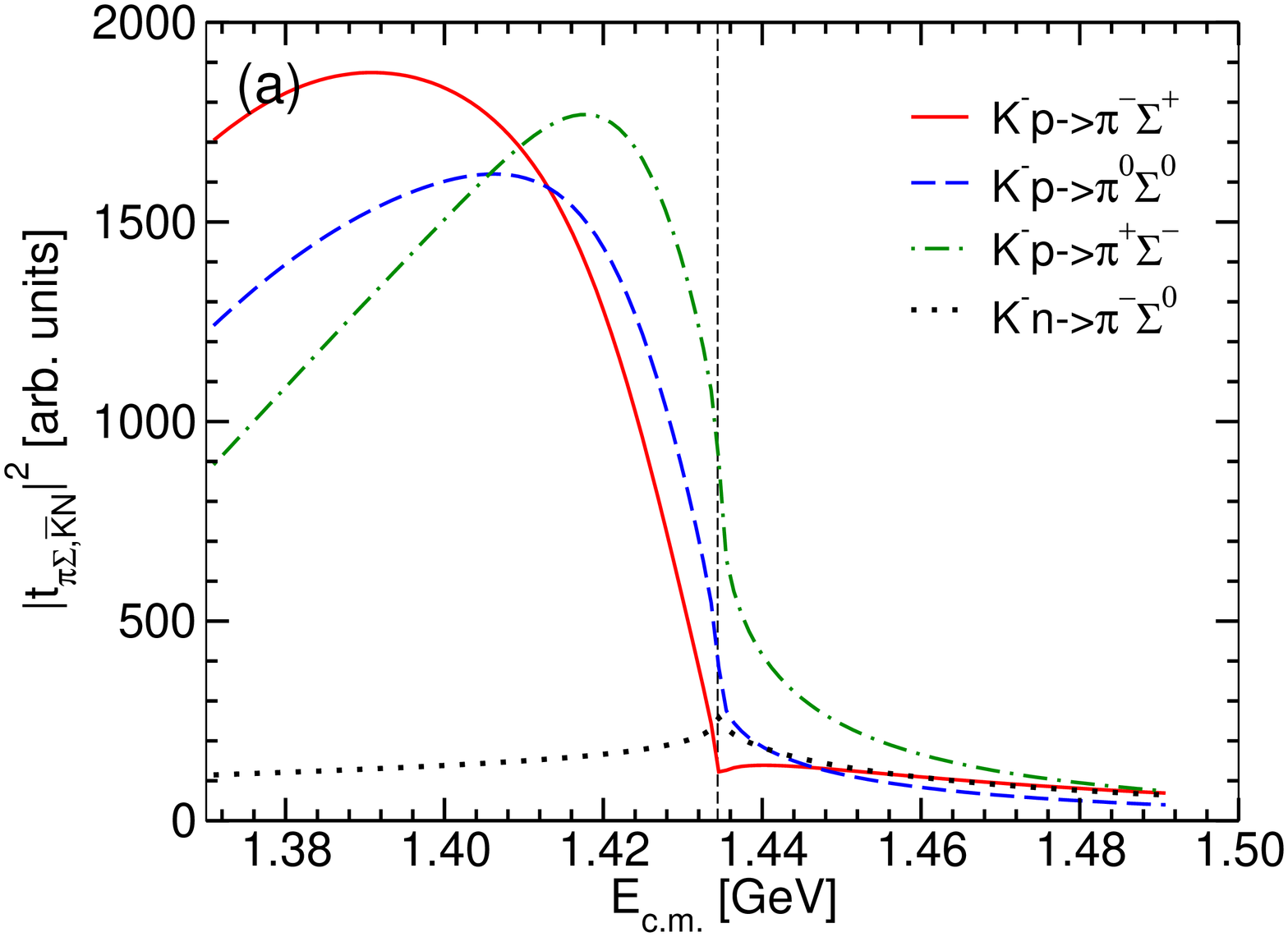}\includegraphics[width=7cm,angle=-90]{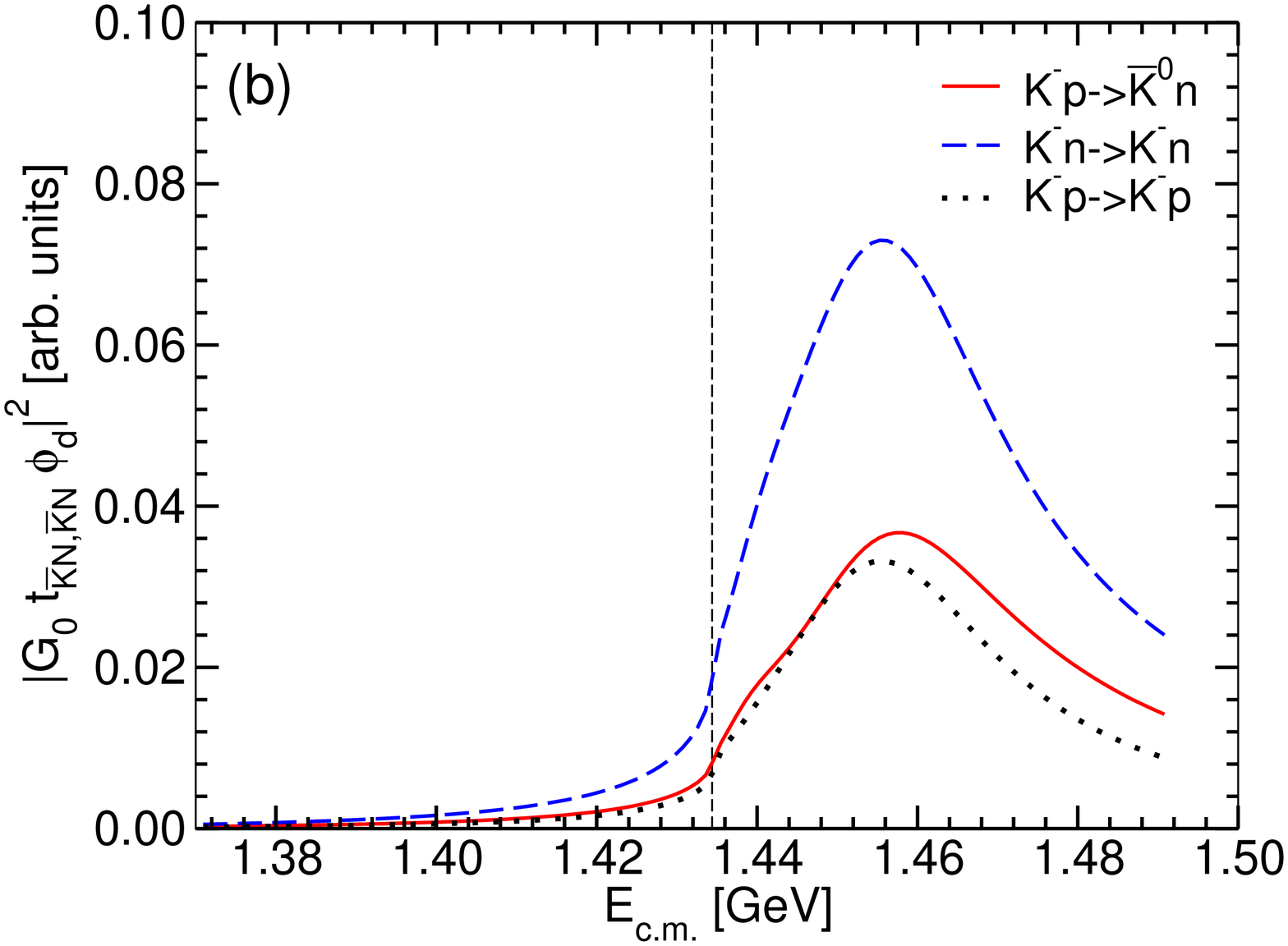}
\caption{ The quantitities (a) $|t_{\pi \Sigma, \bar K N}|^2$ 
and (b) $|G_0\, t_{\bar K N,\bar K N}\, \phi_d|^2$ for different charge channels. 
The results shown are for the potential TW1 \cite{Cieply:2012}. 
Vertical line indicate the $\bar K N$ threshold.
}
\label{f:QFS}
\end{figure*}

\subsection{Origin of the peaks: Quasi-free scattering and $\Lambda$(1405)}

Before analyzing the results in detail and examining also other potentials let
us discuss the origin of the structure of the line shape and, in particular, 
of the peaks. 
It can be understood by considering the amplitude for the two-step process, 
$t_{\pi \Sigma, \bar K N} \, G_0\, t_{\bar K N, \bar K N}\, \phi_d$ which is 
obtained after the iteration of the Faddeev equations~(\ref{tFa2})-(\ref{tFa3}).
The structure results from an interplay between $t_{\pi \Sigma, \bar K N}$ 
and $G_0\, t_{\bar K N, \bar K N}\, \phi_d$.
We present the moduli of those quantities in Fig.~\ref{f:QFS}. 
The $\bar KN \to \pi\Sigma$ amplitude, shown here for different
charge channels, exhibits a clear peak below the $\bar KN$ threshold which 
reflects the presence of the $\Lambda$(1405) resonance. For $\pi\Sigma$ invariant 
masses above that threshold it becomes smooth and rather small. Note that  
$K^-p\to \pi^0\Sigma^0$ (dashed line) corresponds to a pure ($I=0$) isospin state.
In case of $K^-p\to \pi^+\Sigma^-$ and $K^-p\to \pi^-\Sigma^+$ there is
an interference with the $I=1$ state, with opposite signs, and accordingly
the peak positions are shifted to somewhat higher or lower invariant masses.
Moreover, the behavior at the $\bar KN$ threshold is different in case of
$K^-p\to \pi^-\Sigma^+$ (solid line), i.e. there is cusp and not a rounded step anymore. 
Together with the specific weighting by the term $G_0\, t_{\bar K N, \bar K N}\, \phi_d$,
cf. Fig.~\ref{f:QFS}(b), this causes the distinct differences in the cross sections 
for the $\pi^+\Sigma^-n$ and $\pi^-\Sigma^+n$ channels around the 
$\bar KN$ threshold.  

\begin{figure*}[t]
\includegraphics[width=6.8cm,angle=-90]{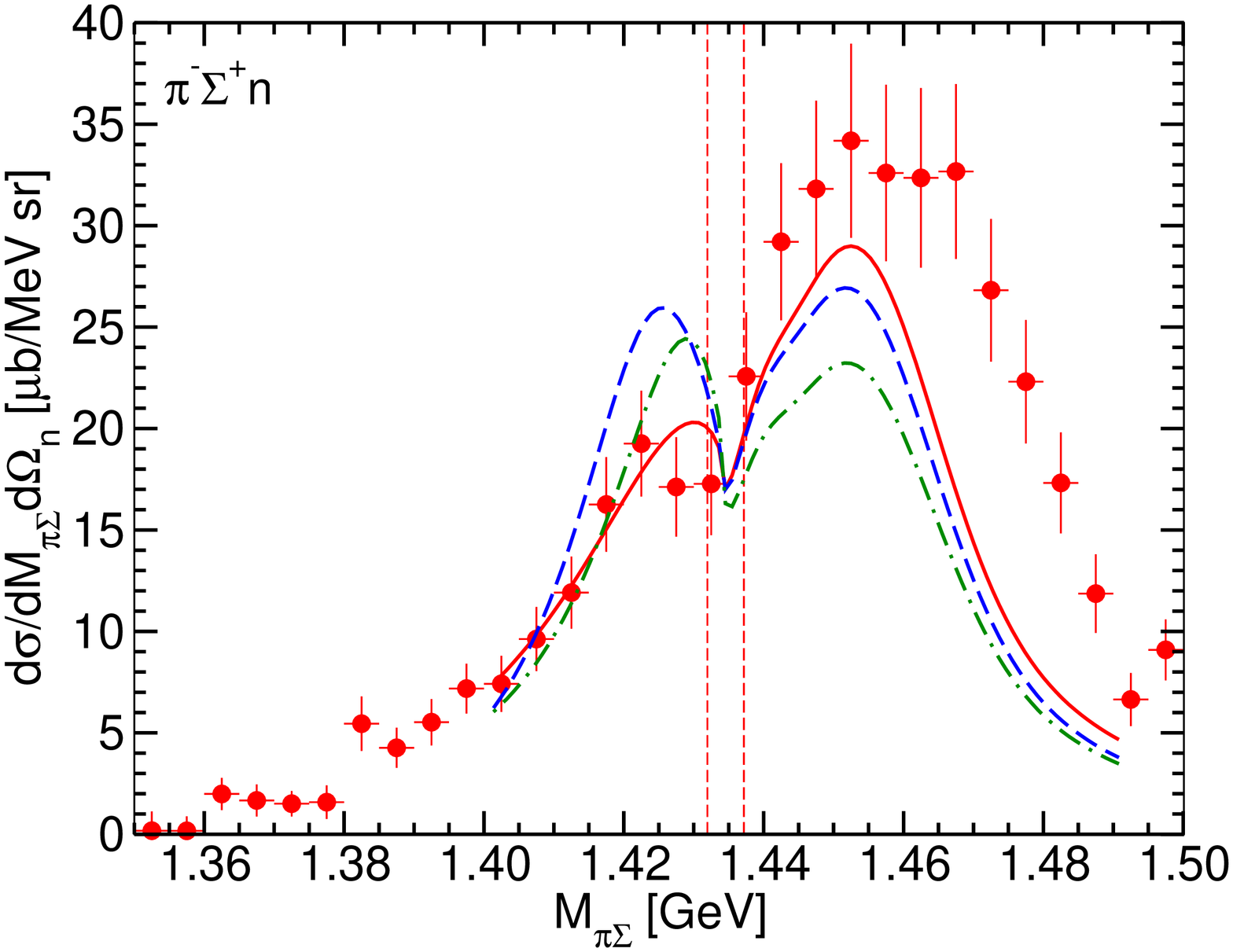}
\includegraphics[width=6.8cm,angle=-90]{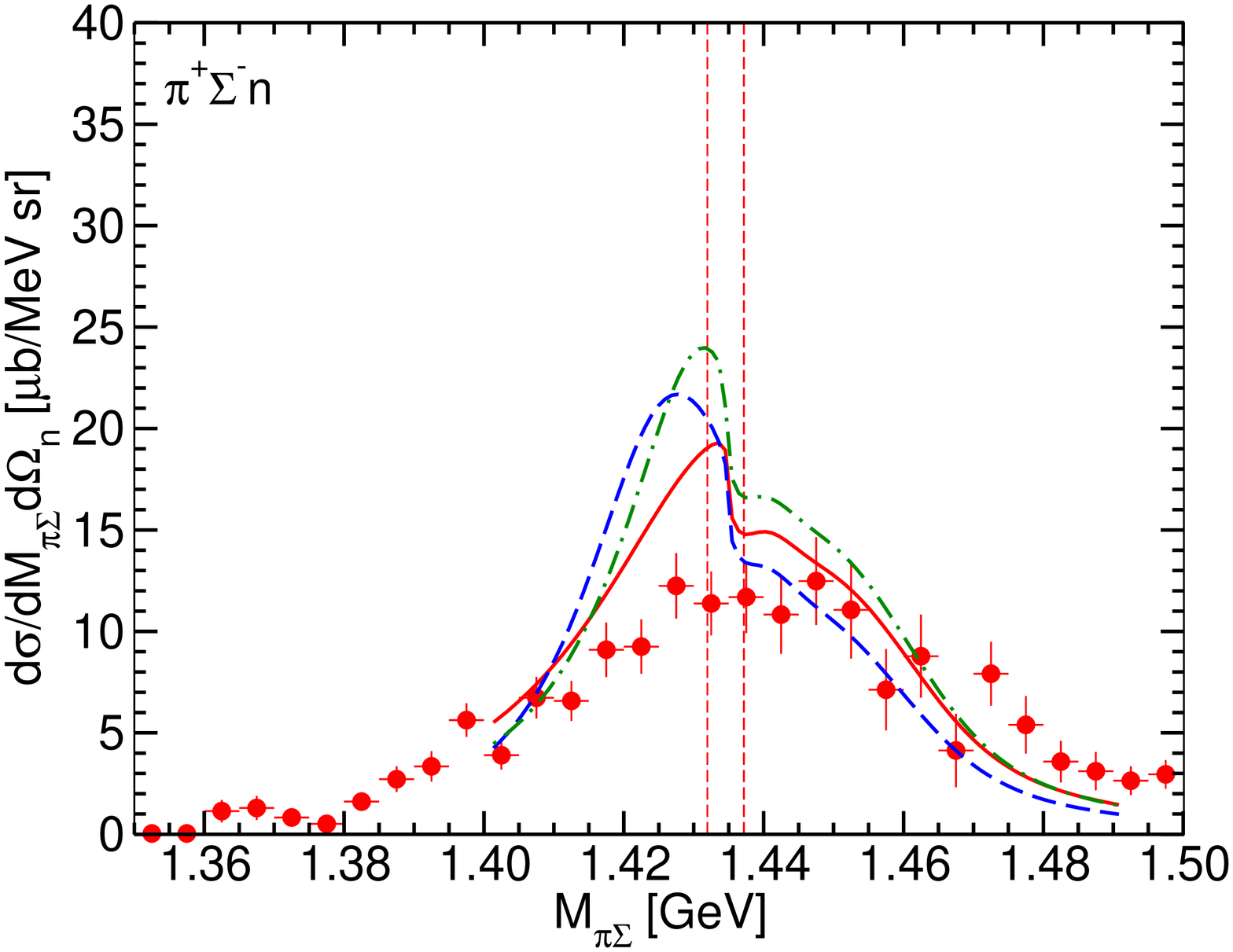}

\includegraphics[width=6.8cm,angle=-90]{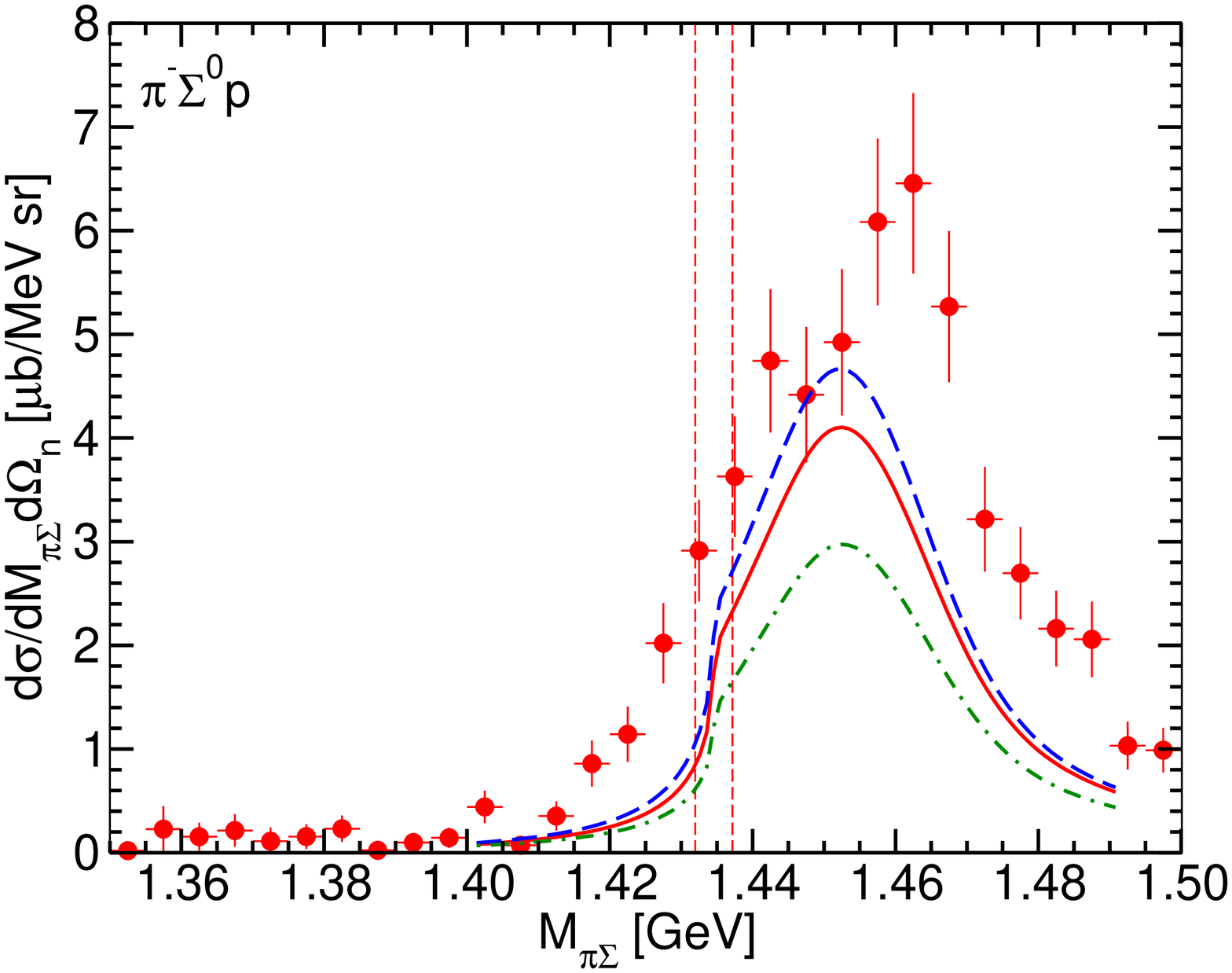}
\includegraphics[width=6.8cm,angle=-90]{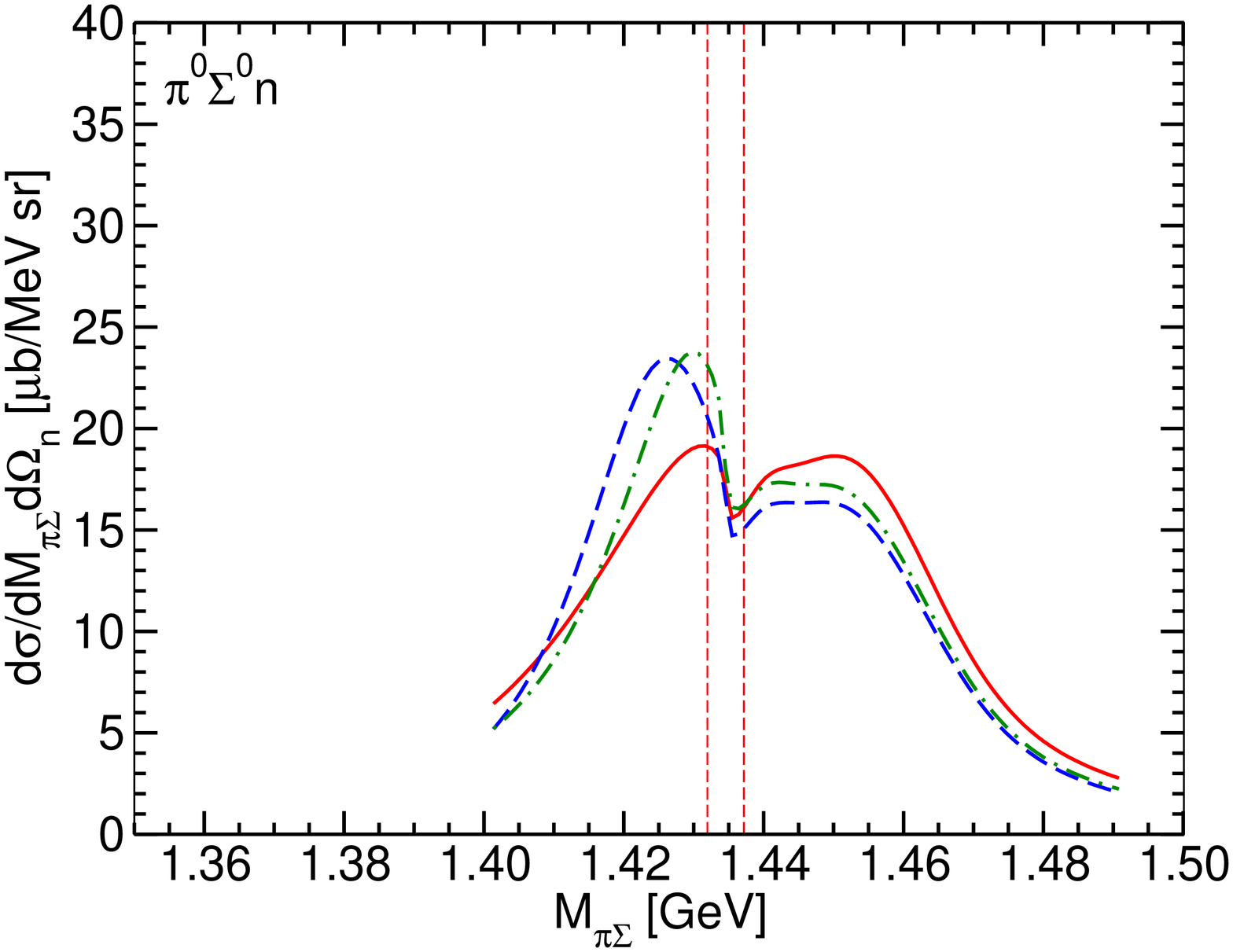}
\caption{Double differential cross section for $K^-d \to \pi\Sigma N$.
Results are shown for the potentials of Ciepl\'y and Smejkal \cite{Cieply:2012}
(solid lines), 
Ohnishi et al.~\cite{Ohnishi:2016} (dash-dotted lines),  
and Oset-Ramos-Bennhold \cite{Oset:2002} (dashed lines). 
Preliminary data are taken from Ref.~\cite{JparcE31A}. 
Vertical lines indicate the $K^-p$ and $\bar K^0 n$ thresholds.
Note that the scale for the $\pi^-\Sigma^0 p$ results is different. 
}
\label{f:XS0}
\end{figure*}

The large peak of the $\pi^-\Sigma^+n$ cross section for energies around $1455$~MeV 
is clearly coming from the combined effect of the Green's function and 
$t_{\bar K N, \bar K N}\,\phi_d$. It is due to quasi-free scattering (QFS) of the $K^-$ on 
the nucleons. The $\pi^+\Sigma^-n$ results are remarkably different,
experimentally as well as in theory, reflecting large interferences between
the $I=0$ and $I=1$ contributions. 
In case of $K^-d\to \pi^-\Sigma^0p$, a pure $I=1$ state, QFS is
likewise responsible for most of the structure, cf. Figs.~\ref{f:QFS}(b) (dotted line) 
and \ref{f:XS1}(b). The $\bar KN \to \pi\Sigma$ amplitude with $I=1$ corresponds to 
the dotted line in Fig.~\ref{f:QFS}(a). There is a noticeable cusp at the $\bar KN$ 
threshold, and the drop off of the amplitude below the threshold is partly responsible 
for the clear reduction of the $K^-d\to \pi^-\Sigma^0p$ cross section in that
energy region, but otherwise $t^{I=1}_{\pi\Sigma,\bar KN}$
exhibits a rather smooth behavior. 

\subsection{Sensitivity to differences between the chiral potentials}

Given that the data from the E31 experiment are still preliminary it is certainly
too early for drawing more detailed conclusions. 
This should be kept in mind when we now compare the predictions based on different 
chiral potentials with each other and confront them also with those data. 
We consider here, besides the potential TW1, the interactions by Ohnishi et 
al.~\cite{Ohnishi:2016} and by Oset-Ramos-Bennhold~\cite{Oset:2002}. Results are 
presented in Fig.~\ref{f:XS0}.

As already discussed above, in our calculation based on TW1 the overall magnitude 
of the cross sections is well reproduced, 
for the $\pi^\pm\Sigma^\mp n$ channels as well as for $\pi^-\Sigma^0 p$. 
(Data for $\pi^0\Sigma^0 n$ have been already presented too \cite{JparcE31A}, but are 
still very preliminary and no absolute values are given.) 
This is also the case for the two other potentials. At the same time, there are 
noticeable variations between the predictions for the different potentials. Since 
the same $\bar KN \to \bar KN$ amplitude is used in the calculations these reflect 
differences in the pertinent $\bar KN \to \pi\Sigma$ amplitudes. 
 
Let us first discuss the $\pi^-\Sigma^0 p$ spectrum which exhibits the simplest structure.
In this channel only the $I=1$ component of $\pi\Sigma$ can contribute and, therefore, one 
can trace back the features in the cross section directly to those of the two ingredients,
$t^{I=1}_{\pi \Sigma, \bar K N}$ and 
$G_0\, t_{K^-p, K^-p}\, \phi_d$,
shown in Figs.~\ref{f:CMP} and~\ref{f:QFS}, respectively. Specifically, there is
a one-to-one correspondence of the order of the maxima in the cross sections and
the magnitudes of the $\bar K N \to \pi \Sigma$ amplitude in the relevant region of 
$1440-1460$ MeV. The ORB potential provides the largest predictions for both. 
 
The $\pi\Sigma$ invariant mass where the maximum occurs is basically determined by the 
product of the Green's function with 
$t_{\bar KN, \bar KN}\, \phi_d$ (cf. Fig.~\ref{f:QFS}(b)) 
and, therefore, it is the same for all potentials. Indeed, also the maximum of the 
two model predictions in Ref.~\cite{Kamano-Lee} is practically at the same invariant mass. 
It is somewhat surprising that the preliminary data suggest that the maximum could be at 
somewhat higher invariant masses. A shift of the maximum by $10-15$ MeV would be rather 
difficult to achieve within our scheme. It would require a drastic change
in the $\bar KN \to\bar KN $ or $\bar KN\to\pi\Sigma $ amplitudes.
Certainly, one has to wait for the final analysis of the experiment in order to 
see whether this discrepancy will persist. 

Interestingly, the $\pi^-\Sigma^0 p$ data do not show any effect from the opening of the 
$\bar KN$ channel.  In the theoretical predictions there is a clear drop off in the cross 
section right below the $\bar KN$ threshold. As a consequence, there is a sizable 
underestimation of the preliminary data in that invariant-mass region.

The $\pi^-\Sigma^+ n$ and $\pi^+\Sigma^- n$ channels involve contributions from $I=0$ and
$I=1$. The line shape for the former is quite well described by the predictions based on 
the potential TW1. This concerns not only the maximum but also the structure 
induced by the $\Lambda$(1405) resonance. Specifically, the calculation produces a
moderate peak at the corresponding invariant mass and a dip at the $\bar K N$ threshold, 
i.e. features that are reasonably well in line with the measurement. Only the peak position 
itself appears to be slightly closer to the $\bar KN$ threshold than what is indicated by the 
preliminary data. Obviously, the other two potential generate a too pronounced structure so 
that the empirical information is drastically overestimated.
Like for $\pi^-\Sigma^0 p$ there is also a noticeable underestimation of the empirical 
results for the $\pi^-\Sigma^+ n$ channel at higher invariant masses for all potentials. 
However, here the available empirical information points to a possible extended plateau rather 
than to an actual shift of the maximum as compared to the theoretical predictions. 

In contrast to the channels discussed above, there is only a poor overall agreement with 
the preliminary data for $K^-d\to \pi^+\Sigma^- n$. 
Here the predictions are only roughly in line with the experiment for higher invariant masses.
Around the $\bar KN$ threshold the spectrum is significantly overestimated. Moreover, 
the structure produced in the three-body calculation does not resemble at all the behaviour
exhibited by the data. In the experiment there is practically no effect seen from the
opening of the $\bar KN$ channel, while theory produces a pronounced peak below
the threshold for all considered potentials. Actually, the same incorrect behavior 
is present in the results by Kamano and Lee~\cite{Kamano-Lee}. 

Finally, the results for $\pi^0\Sigma^0 n$ are qualitively similar to those for
$\pi^-\Sigma^+ n$, except that there is 
a less pronounced maximum for invariant masses
above the $\bar KN$ threshold. In this channel only the $I=0$ component of $\pi\Sigma$ 
can contribute so that it is the most promising one for exploring and pinning
down the structure of the $\Lambda$(1405) resonance. 
 
\begin{figure}
\includegraphics[width=7cm,angle=-90]{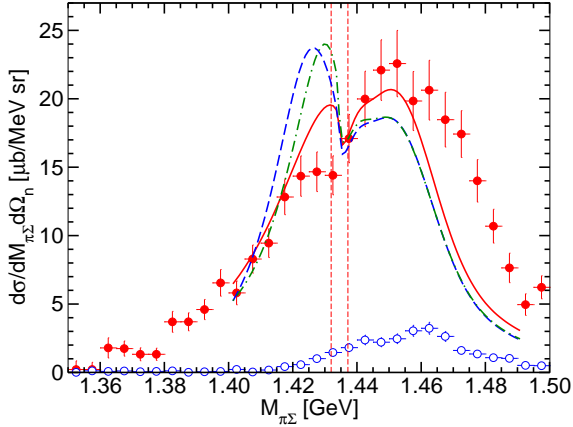}
\caption{Average of $K^-d\to\pi^+\Sigma^- n$ and
$K^-d\to\pi^-\Sigma^+ n$ spectrum (lines, filled circles) vs. 
half of the $K^-d\to\pi^-\Sigma^0 p$ spectrum (empty circles), 
cf. Eq.~(\ref{XS}) and text. 
Same description of curves as in Fig.~\ref{f:XS0}. 
Prelimary data are taken from Ref.~\cite{JparcE31A}. 
Vertical lines indicate the $K^-p$ and $\bar K^0 n$ thresholds.
}
\label{f:XSM1}
\end{figure}

Following the experimentalists we consider here in addition the average of the 
$K^-d\to\pi^+\Sigma^- n$ and $K^-d\to\pi^-\Sigma^+ n$ spectrum, 
cf. Fig.~\ref{f:XSM1}, denoted by  
$(\sigma_{\pi^+\Sigma^- n} + \sigma_{\pi^-\Sigma^+ n})/2$ to simplify the notation. 
Since \cite{Hyodo:2012} 
\begin{equation}
\sigma_{\pi^\pm\Sigma^\mp n} \propto 
\frac{1}{3}|T_{I=0}|^2 + \frac{1}{2}|T_{I=1}|^2 \pm
\frac{\sqrt{6}}{3}{\rm Re}\,(T_{I=0}T^*_{I=1}) \ ,
\label{XS} 
\end{equation}
it is clear that in this average the interference term between the $I=0$ and $I=1$ 
contributions drops out.  
We want to emphasize, however, that the amplitudes $T_{I}$ in Eq.~(\ref{XS}) do 
not correspond directly to those for $\bar K N\to \pi\Sigma$, i.e. to 
$t^I_{\pi\Sigma,\bar K N}$.
Formally, and ignoring the interdependence of the kinematical variables for which 
the amplitudes in the subsystems are evaluated, their relations are 
$T_{I=0} = t^{I=0}_{\pi\Sigma,\bar K N}/\sqrt{2} \times (A-B)$ and
$T_{I=1} = t^{I=1}_{\pi\Sigma,\bar K N}/\sqrt{2} \times (A+B)$, where 
$A = G_0\,t_{K^0 n,K^-p} \phi_d$ and    
$B = G_0\,t_{ K^- n,K^-n} \phi_d$. 
Here the relative signs reflecting whether the proton or neutron in the deuteron
takes part in the scattering process have been already accounted for. 
 
Furthermore, one should be aware that due to the large mass splitting
between $K^-$ and $\bar K^0$ \cite{PDG}, the physical thresholds of 
the $K^-p$ and $\bar K^0 n$ channels are separated by more than $5$~MeV,
as indicated in Fig.~\ref{f:XSM1}. 
Thus, there will be a potentially large breaking of the isospin 
symmetry in the region close to and between the $K^-p$ and $\bar K^0 n$ thresholds  
making it impossible to define amplitudes with proper isospin. 
Consequently, caution is required when applying Eq.~(\ref{XS}) 
for the interpretation of the data in that specific energy region. 

Fig.~\ref{f:XSM1} includes also data for (half of) $\sigma_{\pi^-\Sigma^0 p}$ (empty circles).
Since that cross section corresponds to $\frac{1}{2}|T_{I=1}|^2$ it is
obvious that the $\pi^\pm\Sigma^\mp n$ results are dominated by the $I=0$ component. 
Nevertheless, the individual results shown in Fig.~\ref{f:XS0} reveal that 
the $I=1$ contribution is by no means negligible and plays a decisive 
role for the actual line shapes. 

In view of the preliminary character of the data it is premature to draw more 
concrete conclusions with regard to the properties of the elementary 
$\bar KN\to \pi\Sigma$ interaction. However, it is obvious that larger values
of the $I=1$ $\bar KN\to \pi\Sigma$ amplitude for invariant masses above
the $\bar K N$ threshold would bring the maximum of the $\pi^-\Sigma^0 p$
cross section closer to the experiment and likely also the one for
$\pi^-\Sigma^+n$. Indeed, the KSU analysis supports larger values for
the corresponding $s$-wave amplitude, see Fig.~\ref{f:CMP}. Its absolute square 
exceeds the one predicted, e.g., by the ORB potential by about 30~\% at $1480$ MeV. 

The situation is more complicated below the $\bar K N$ threshold and,
specifically, in the $\Lambda$(1405) region. Still, the results shown in 
Figs.~\ref{f:XS0} and \ref{f:XSM1} provide a strong indication that the 
$\Lambda$(1405) peaks by all three potentials are too large in magnitude.
In particular, there is a dramatic overestimation in the sum of $\pi^+\Sigma^-n$ 
and $\pi^-\Sigma^+n$, cf. Fig.~\ref{f:XSM1}, where interferences between the 
$I=0$ and $I=1$ amplitudes should cancel, at least to some extent and
disregarding the potential difficulties with the isospin ``interpretation''
mentioned above. 

\begin{figure}
\includegraphics[width=7cm,angle=00,clip=true]{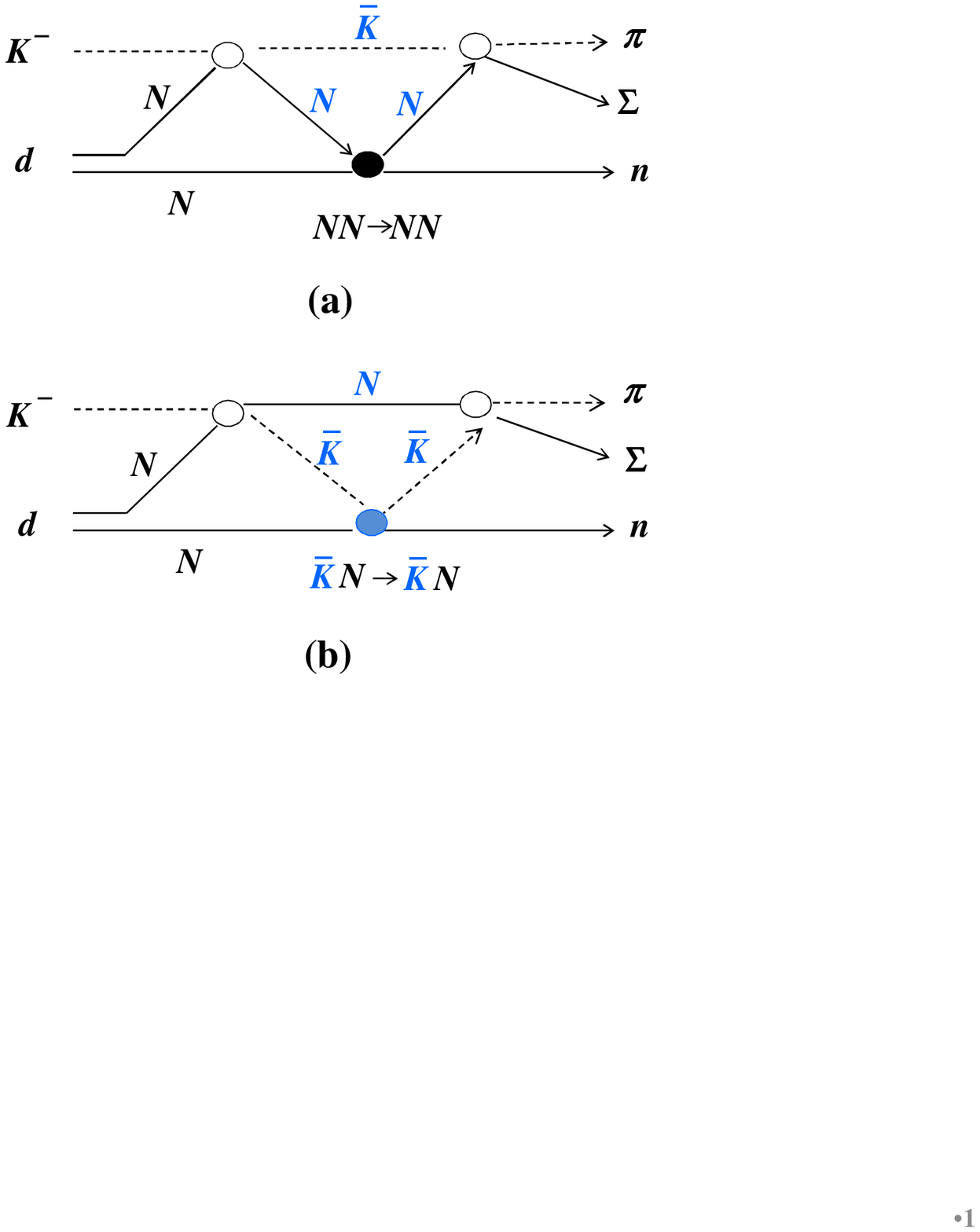}
\caption{Diagrams for three-step processes 
}
\label{f:3step}
\end{figure}

\subsection{Influence of 3-step processes}

\begin{figure}
\includegraphics[width=7cm,angle=-90]{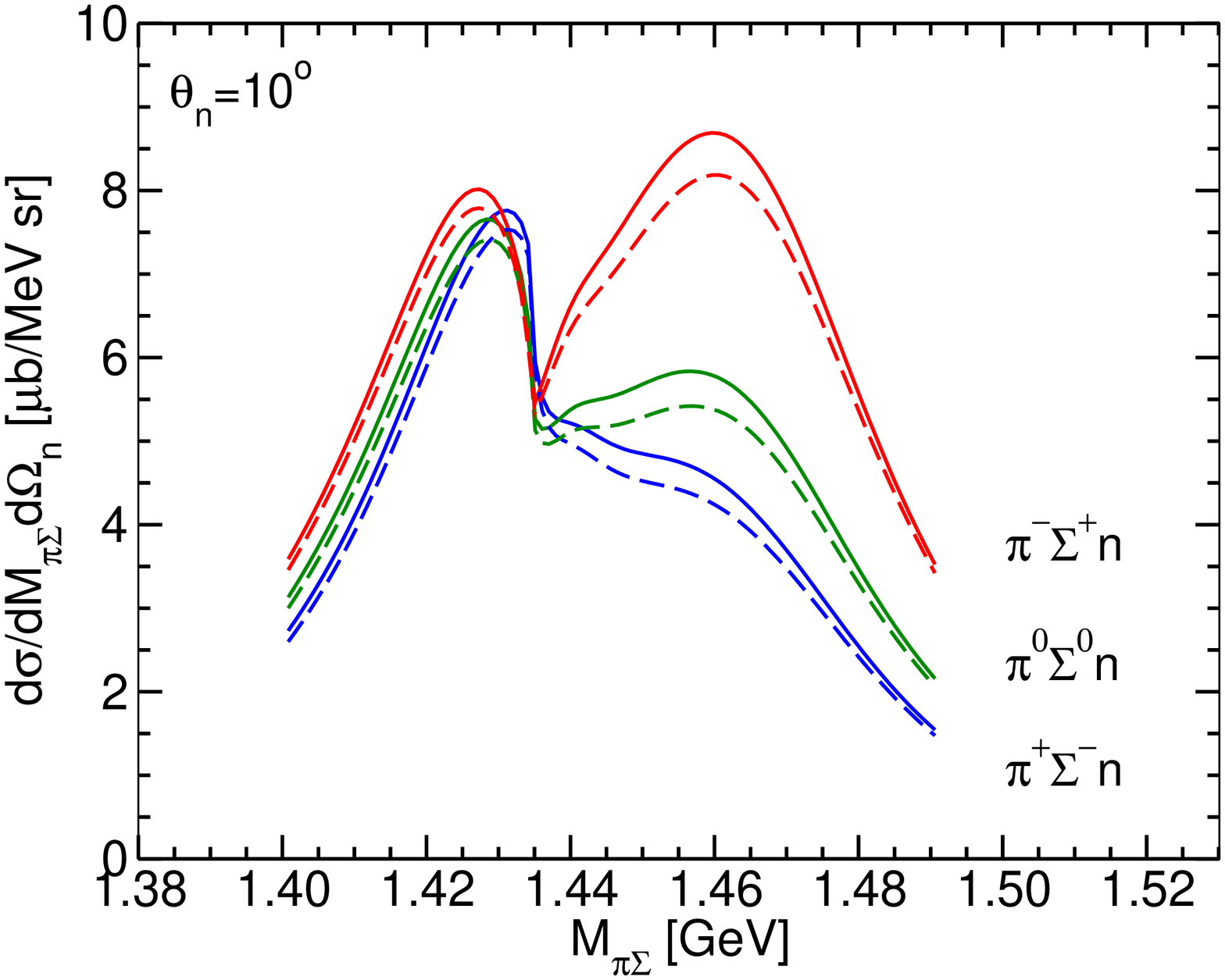}

\includegraphics[width=7cm,angle=-90]{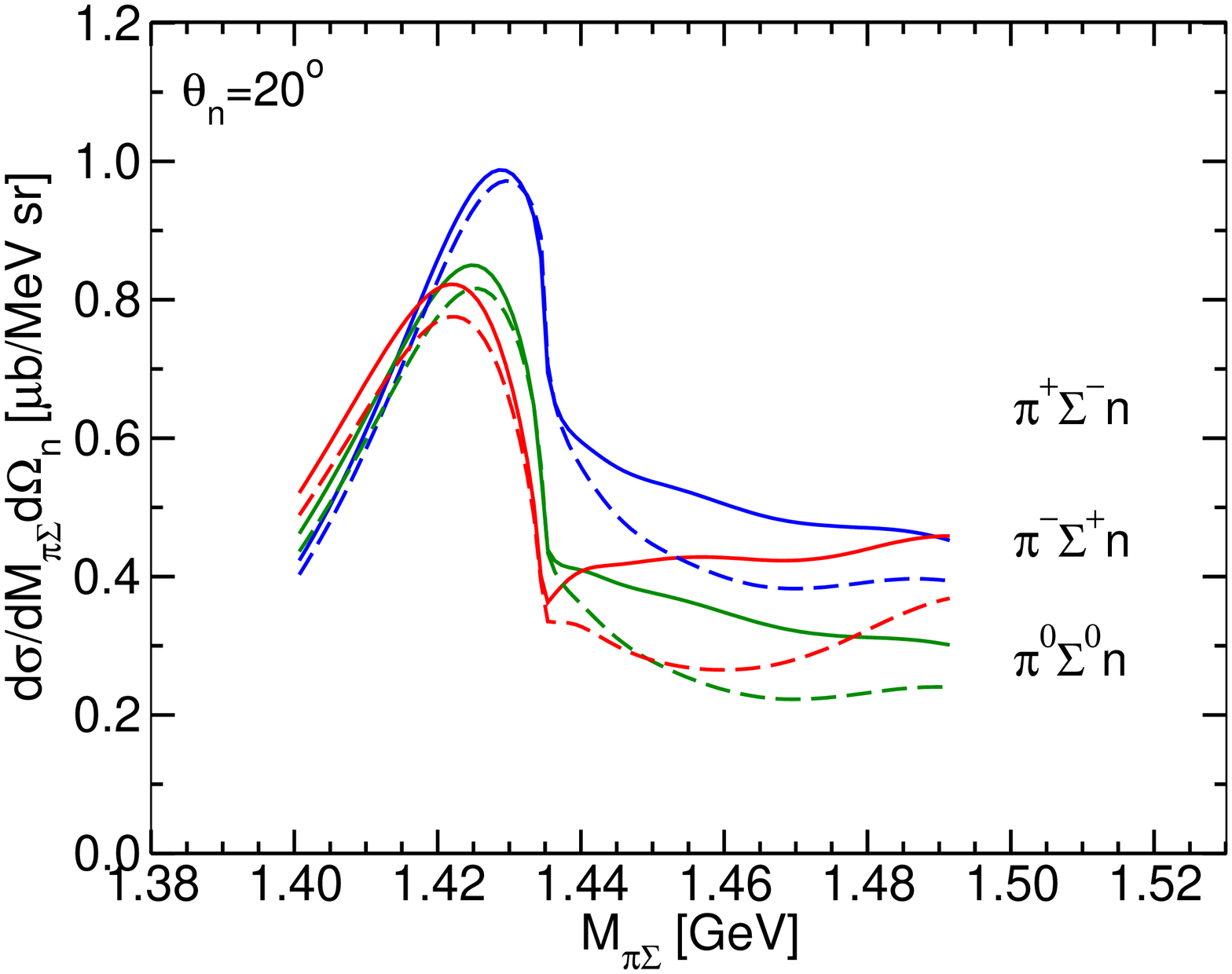}
\caption{$K^-d\to\pi\Sigma n$ results for neutron angles  
$\theta_n=10^\circ$ and $20^\circ$. 
Results based on two-step processes only (solid lines) are shown together with those
where three-step processes are also included (dashed lines). 
The predictions are based on the potential TW1 \cite{Cieply:2012}. 
}
\label{f:XS3st}
\end{figure}

In the course of our calculation we explored also contributions from three-step 
processes, where the corresponding amplitudes are obtained by iterating twice 
the Faddeev equations~(\ref{tFa1})-(\ref{tFa5}). As already mentioned
we needed  several iterations to reach converged results in case when only the $s$-wave 
$\bar KN \to \bar KN$ amplitude was included~\cite{HYP2015}.
In contrast, now where $\bar KN$ partial waves up to $j=7/2$ are incorporated, 
there are practically no visible changes in the invariant-mass spectra at $0$ degree of the 
outgoing neutron when the three-step processes are included. 
Let us provide exemplary results for the two processes shown in Fig.~\ref{f:3step} which are 
expected to yield the largest effects among the three-step processes. 
We use the Nijmegen potential Nijm93~\cite{Nij93NN} for the $NN$ sector, and include partial waves 
up to $j=3$. It turned out that the process Fig.~\ref{f:3step}(b) gives larger contributions than 
the process Fig.~\ref{f:3step}(a) with the $NN$ interaction, but still the overall effect is tiny
for $\theta_n=0^\circ$ and the spectra remain almost unchanged as compared to the results
for the two-step processes. Therefore, we extended the calculations to $\theta_n=10^\circ$ and 
$20^\circ$ for further exploration. Corresponding results including the process 
Fig.~\ref{f:3step}(b) are shown in Fig.~\ref{f:XS3st}.
With increasing neutron angle the peak originating from $\bar KN$ QFS is more and more reduced 
and the structure due to the $\Lambda$(1405) becomes more pronounced.
However, at the same time the overall magnitude of the cross section is strongly reduced,
which makes corresponding experiments much more challenging. 

\section{Summary} 

In this paper we reported on a calculation of the reaction $K^-d \to \pi\Sigma N$
within a Faddeev-type approach. 
The work is motivated by corresponding experiments that are presently performed
at J-PARC. Accordingly, spectra for various charge channels of the 
$\pi\Sigma$ final state are presented for the specific kinematics of the E31 
experiment \cite{Jparc}, namely for the $K^-$ beam momentum of $p_K = 1$~GeV/c and 
the neutron angle of $\theta_n=0^\circ$. A comparison with preliminary data 
that have become available recently \cite{JparcE31,JparcE31A} was performed.  

The employed Faddeev-type approach requires as main input amplitudes for the
two-body subsystems $\bar KN \to \bar KN$ and $\bar KN \to \pi\Sigma$. 
For the latter we utilized results generated from so-called chiral unitary models
taken from the literature. Specifically, we used the potentials by Ciepl\'y and 
Smejkal (TW1) \cite{Cieply:2012} and Ohnishi at al. ($V^{E-dep.}$)
\cite{Ohnishi:2016}, that both are constrained by the latest measurement of
kaonic hydrogen \cite{Bazzi:2011}, and a more historical potential that is due 
to Oset, Ramos and Bennhold \cite{Oset:2002}.   
On the other hand, the $\bar KN \to \bar KN$ amplitude was taken from a recent 
partial-wave analysis \cite{KSU} -- because of the following reason:  
While in the calculation of the quantities measured in the E31 experiment, 
the $\bar KN \to \pi\Sigma$ amplitude is sampled at energies around the 
$\bar K N$ threshold, the $\bar KN \to \bar KN$ amplitude is probed in an entirely
different kinematical regime. It is required for c.m. energies corresponding 
to the initial momentum of $p_K = 1$~GeV/c, which means around $1800$~MeV. 
The aforementioned chiral potentials do not provide a realistic description
of $\bar KN$ scattering at such high energies. Moreover, in that energy region
higher partial waves yield an essential contribution, not only for $\bar KN$
elastic scattering but also in the reaction $K^-d \to \pi\Sigma N$ that is
investigated here. The latter aspect has become clear after the pioneering 
work of Kamano and Lee \cite{Kamano-Lee}, and it has been confirmed in the 
present study. Chiral potentials are typically limited to $s$-waves. 
Thus, calculations that employ such models for the $\bar KN \to \pi\Sigma$ as
well as the $\bar KN \to \bar KN$ amplitudes  - like ours \cite{KM:2012} and 
several others in the past - allow only very limited access to the physics 
that governs the E31 experiment.

The predictions of our calculations turned out to agree quite well with the 
preliminary data on a qualitative level, i.e. as far as the magnitude and the
line shape in general is concerned. Especially, the spectra for 
$K^-d \to \pi^-\Sigma^+n$ and $K^-d \to \pi^-\Sigma^0p$ are fairly close to
the data reported so far. However, on a more quantitative level there are
noticeable differences. In particular, the situation with regard to the
structure of the $\Lambda$(1405) -- the prime motivation behind the E31 experiment --
is conflicting. Indeed, all three potentials produce a structure in the
relevant $\pi\Sigma$ invariant-mass region -- however, it is much too
pronounced as compared to what is indicated presently by the measurement. 
Actually, in case of $\pi^+\Sigma^-n$ even the line shape is quite different. 

Given the preliminary status of the data it is obvious that the present study can
only have an exploratory character and solid conclusions, specifically with 
regard to the structure of the $\Lambda$(1405) resonance, have to be postponed. 
Nonetheless, it has become clear that the general conditions are similar to 
what has been already found in studies of other reactions with the aim of scrutinizing 
the structure of the $\Lambda$(1405) \cite{Roca:2013,Nakamura:2013,Mai:2014}, 
namely that the line shape around the $\bar KN$ threshold is a result of 
(a) a delicate interplay between the isospin $I=0$ and $I=1$ $\bar K N\to \pi \Sigma$ 
amplitudes, and 
(b) the energy dependence of the sub-threshold $I=0$ $\bar K N$ amplitude 
or, equivalently, the pole structure of the $\Lambda$(1405).
Disentangling these two aspects remains a challenge. 
In any case the observed differences between the employed potentials  
are promising for the prospect of getting further constraints on the 
$\bar K N$ interaction in the $\Lambda$(1405) resonance region and,
specifically, on the $\bar K N\to \pi\Sigma$ transition amplitude. 
Of course, whether, finally, conclusions on the $\Lambda$(1405) will be possible or not, 
depends not least on the accuracy of the data that is eventually achieved in the
E31 experiment. The most promising channel would be
$K^-d \to \pi^0\Sigma^0 n$, where the emerging $\pi^0\Sigma^0$ system is in 
a pure isospin $I=0$ state. However, with only neutral particles in the final state 
it is obviously also by far the most ambitious one for an experiment  
\cite{JparcE31A}.

\acknowledgments{
We acknowledge communication with A.~Ciepl\'y and M.~Mai with regard 
to their $\bar KN$ interactions.
}

\appendix
\section{Permutation operator}
\label{permutation}
We use the Balian-Br\'ezin \cite{Polyzou,BB1} approach to calculate the permutation matrix element
in Eq.~(\ref{P23}).
Since the details of the derivation  for the non-relativistic case are given in Ref.~\cite{Newform},
we concentrate on  an extension to the relativistic case and  only show its final expression.
In Sect.~\ref{sec:tec1}, 
$\braket{  k'  q\, \alpha' | P_{23} | \, k'' q'' \alpha'' }$
appears, but 
a cyclic permutation
$ P_{23}P_{13} $ often used  is considered
because
$\bra{  k'  q\, \alpha'}P_{23} $
is easily obtained from 
$\bra{  k'  q\, \alpha'} P_{23}P_{13}$ by a permutation inside the two-body sector.

First, we introduce the momentum state of the non-interacting particles 1 and 2 in the two-body c.m. frame 
which is associated with  individual momenta   
via 
$$\ket{ \vec{k}''; \vec{0}}\equiv    \ket{\vec{k}''}\ket{- \vec{k}''}  \: . $$
On the left-hand side, the relative momentum $\vec{k}''$ and the total momentum zero are
indicated.   
This is boosted to the three-body c.m. frame and expressed together with the third particle (numbered 3) state
$\ket{\vec{q}\, ''}$  as
$$\ket{ \vec{k}''; -\vec{q}\, ''} \ket{\vec{q}\, '' }$$   
where the Wigner rotations for spins are neglected. 
The permutation matrix element between  these states
can be evaluated by using Eq.~(\ref{Eqfong}) as
\begin{align}
& \bra{ \vec{k}'; -\vec{q\,} } \bra{\vec{q\,}}
  \,  P_{23}P_{13}\,  \ket{ \vec{k}''; -\vec{q}\,''} \ket{\vec{q}\,'' }  \nonumber \\
&=\delta (\vec{k}'-\rho\, \vec{q}-\vec{q}\,'' )\: \delta (\vec{k}''+ \vec{q}+\rho''\vec{q}\,'' )
\: \delta (\vec{q}+ \vec{q}\,''+\vec{q}_1 ) \nonumber\\
&\:\:\times N^{\, -\frac{1}{2}} N''^{\, -\frac{1}{2}}
 \label{vecP} 
\end{align}
where
\begin{eqnarray}
N=
\Bigg| \frac{\partial (\vec{q''},\vec{q_1})}  {\partial (-\vec{q}, \vec{k'})}\Bigg|  
=\frac{W_{31}}{\omega_{31}} \, \frac{\omega_3(q'')\; \omega_1(q_1)} {u_3(k')\; u_1(k')}
\nonumber
\end{eqnarray}
is the Jacobian \cite{relativisticND} for the Lorentz transformation from $(\vec{q}\,'' ,\vec{q}_1)$ 
to  $(-\vec{q}, \vec{k}' )$.  Similarlily,
\begin{eqnarray}
N''=
 \Bigg| \frac{\partial (\vec{q}_1,\vec{q})}  {\partial (-\vec{q}\,'',\vec{k}'')}\Bigg|
=\frac{W_{12}}{\omega_{12}}\,  \frac{\omega_1(q_1)\; \omega_2(q)} {u_1(k'')\; u_2(k'')}
\nonumber
\\
\nonumber
\end{eqnarray}
\noindent
where $\omega_i(q)=\sqrt{q^2+m_i^2}$, $u_i(k')=\sqrt{k'^2+m_i^2}$, $W_{31}=u_3(k')+u_1(k')$,
$W_{12}=u_1(k'')+u_2(k'')$  and $\omega_{ij}=\omega_i+\omega_j$ for 
{\scalebox{0.9}{$\displaystyle i,j=1,2,3$}}.

Equation~ (\ref{vecP}) is of similar form as the one defined with Jacobi momenta in the non-relativistic case,
except for  the Jacobians, but $\rho$ and $\rho''$ are no longer constants.
Those are expressed as 
\begin{eqnarray}
\rho =\frac{1}{W_{31}} \Big\{ \frac{ -\vec{k}' \cdot \vec{q}  }  { \omega_{31}+W_{31} } +u_3 (k') \Big\}
\label{eqRho}
\end{eqnarray}
and
\begin{eqnarray}
\rho'' =\frac{1}{W_{12}} \Big\{ \frac{ \vec{q} \cdot \vec{q}\,'' }  { \omega_{12}+W_{12} } +\omega_2 (q) \Big\} \: . 
\end{eqnarray}
We do not want to go into further details in this paper, but mention that the Jacobians  $N$, $N''$ and $\rho$, $\rho''$
are expressed by only three variables $k'$, $q$ and $x$, where $x$ is defined as $x\equiv \hat k'\cdot \hat q$.

On the basis of the above results, the permutation matrix element  between partial-wave projected basis
states
$\braket{  k'  q\, \alpha | P_{23} P_{13} | \, k'' q'' \alpha' }$
can be  evaluated in line with Appendix A  in \cite{Newform}.  The resulting form is
\begin{align}
 &\braket{  k'  q\, \alpha \, |\, P_{23} P_{13} | \, k'' q'' \alpha' } = 
 \nonumber\\
 &\int_{-1}^{1} dx \,  \frac{ \delta (q''-\chi )} {q''^2} \,  \frac{ \delta (k''-\pi )} {k''^2} \bar R_{\alpha\alpha'}(k' q\, x) \, ,
\end{align} 
where $\chi$ and $\pi$ are given in Eqs.~(\ref{chi}), (\ref{pi}), and
\begin{align}
&\bar R_{\alpha\alpha'}(k' q\, x) = N^{\, -\frac{1}{2}} N''^{\, -\frac{1}{2}}
\nonumber \\
&\times \sqrt{\hat j \, \hat j_s\, \hat j' \,\hat j_s'}   \sum_{L S} \hat S
  \left\{
  \begin{array}{ccc}
        l & s & j \\
       \lambda &  s_2 & j_s \\
       L & S & J \\
  \end{array}
  \right\}
  \left\{
  \begin{array}{ccc}
        l' & s' & j' \\
       \lambda' &  s_3 & j_s' \\
       L & S & J \\
  \end{array}
   \right\}
  \nonumber\\
& \times (-)^{s_2+2 s_3+s'} \sqrt{\hat s \hat s'}
  \left\{
  \begin{array}{ccc}
        s_3 & 0 & s  \\
       s_2 &  S  & s' \\
  \end{array}
   \right\}
  \nonumber\\   
& \times 8 \pi^2 \sum_{m_l\, m_{l'}\, m_{\lambda'}}
  (l\, \lambda\, L,\,m_l\, 0\, m_l)\:  (l'\, \lambda'\, L,\,m_{l'}\, m_{\lambda'}\, m_l)
  \nonumber\\  
& \times(-)^{m_l} Y_{l\, -m_l}(\hat k') Y_{l'\, m_{l'}}(\hat k'') Y_{\lambda' \, m_{\lambda'}}(\hat q'')
{\scalebox{0.85}{$\displaystyle\sqrt{  \frac{2 l+1}{4\pi} } $}}   \:\: .  
\nonumber\\   
\end{align}
We use the notation $\hat j\equiv 2j+1$, and  assume that
$\vec{q}$ is along the $z$-axis and $\vec{k}'$ lies in the 
$x$-$z$ plane. The two vectors $\vec k''$ and $\vec q\,''$ are 
defined as
\begin{align}
&\vec q\, ''=\vec k'-\rho\, \vec q  \:\:,  \nonumber\\
&\vec k''=-\rho''\vec k'-(1-\rho''\rho)\,\vec q \:\: .
\nonumber
\end{align}

Finally,  $R_{\alpha\alpha'}$ 
in Eq.~(\ref{P23}) is 
related to  $\bar R_{\alpha\alpha'}$ 
by
$$R_{\alpha\alpha'}(k' q\, x)=(-)^l \bar R_{\alpha\alpha'}(k' q\, x) \, , $$ 
where the phase is easily obtained  by applying $P_{13}$ on
 $\bra{  k'  q\, \alpha }P_{23} $ to the left as mentioned above. 

\section{Analytical integration over $x$ and  $k''$  and the domain for the $k'$ and $q''$ integrations} 
\label{integration:x}
Here we describe how the $x$ and  $k''$ integrations are performed analytically in Eq.~(\ref{T12}) and how
the domain for the $k'$ and $q''$ integrations is defined in Eq.~(\ref{T12f}). 
The major advantage of choosing $x$ and $k''$ variables for analytic integrations 
is that it enables us to avoid  moving singularities  which are well known to be difficult to treat in three-body calculations.
This prescription was presented in Sec. 2.2 of \cite{Newform} for the non-relativistic case which is
somewhat simpler. 
We explain how to extend it to the relativistic case and show only the formulas without 
going into details  with regard to the numerical calculations.

In order to rewrite $\delta (q''-\chi )$ in Eq.~(\ref{P23})  in the form where $x$ is explicitly shown,
first we  deform $\rho$ given in Eq.~(\ref{eqRho}) as follows:
\begin{eqnarray}
\rho =\sigma (1-\delta x)
\label{eq:rho}
\end{eqnarray}
where
\begin{eqnarray}
\sigma=\frac{u_3(k')}{W_{31}}\: , \hskip10pt 
\delta=\frac{k'q}{u_3(k')(\omega_{31}+W_{31})} \:\: .
\nonumber
\end{eqnarray}
Notice that $\sigma$ and $\delta$ are functions of $k'$ and $q$. Then
$\delta (q''-\chi )$ is rewritten as
\begin{eqnarray}
 \delta (q''-\chi ) =\frac{  q'' } {k'\rho\, q f_r } \, \delta (x-x_0) \, \Theta (1-|x_0|) \,  ,
\label{eq:delx0}
\end{eqnarray}
where 
\begin{eqnarray}
 f_r =\Big| 1 +\frac{\sigma q}{ k'}\,\delta   - \frac{\delta \, x_0}{1-\delta \, x_0} \Big| \, , 
\nonumber
\end{eqnarray}
and $x_0$ is a solution of 
\begin{eqnarray}
   q''^2= k'^2+\rho^2 q^2 -2k'\rho\, q x
\label{eq:q''}
\end{eqnarray}  
(see Eq.~(\ref{chi})). Since $\rho$ is a linear function of $x$, Eq.~(\ref{eq:q''}) has actually 
two solutions, but one of them turns out to be physically meaningless.
 We omit the lengthy expression of $x_0$ here, but mention that it is a function of
 $k'$, $q$ and $q''$.
 
 Using the two $\delta$-functions, $\delta (x-x_0)$ and $\delta (k''-\pi)$  we can perform
 the $x$ and $k''$ integration analytically in Eq.~(\ref{T12}). 
Note that the $\Theta$-function in Eq.~(\ref{eq:delx0})  restricts and
 defines the domain for the double integrations over $k'$ and $q''$.
 In the non-relativistic case, $\rho$ is a constant 
 ({\scalebox{0.85}{$\displaystyle \rho=m_3/(m_3+m_1)$}} )
 and  the domain is easily deduced from 
Eq.~(\ref{eq:q''}) and $|x_0|\le 1$. It becomes an open rectangular region in the $k'$-$q''$ plane 
restricted by the three straight lines, $q''=k'-\rho\, q$ and  $q''=\pm k'+\rho\, q$  (see Fig.~1 in \cite{Newform}).
In the relativistic case, $\rho$\, (=$\sigma (1-\delta x_0$) ) depends on $k'$, $q$ and $x_0$, namely
$k'$, $q$ and $q''$, the boundaries of the ``rectangular'' are no longer straight lines but curves. 
Those are given by  
\begin{align} 
&q''= \:\:\, k'+\rho_{\scalebox{0.7}{$\displaystyle -$}} \, q  \:\: , \nonumber\\
&q''= \:\:\,  k'-\rho_{\scalebox{0.7}{$\displaystyle +$}} \, q \:\: , \nonumber\\
&q''=   -k'+\rho_{\scalebox{0.7}{$\displaystyle +$}} \, q \:\: . \nonumber
\end{align}
where $\rho_{\scalebox{0.7}{$\displaystyle -$}}=\sigma (1+\delta)$ and
 $\rho_{\scalebox{0.7}{$\displaystyle +$}}=\sigma (1-\delta)$.
 Thus we arrive at the expressions in  Eq.~(\ref{T12f}).



\begin{thebibliography}{20}

\bibitem{Hyodo:2012} 
  T.~Hyodo and D.~Jido,
  Prog.\ Part.\ Nucl.\ Phys.\  {\bf 67}, 55 (2012). 

\bibitem{Cieply:2016} 
  A.~Ciepl\'y, M.~Mai, U.-G.~Mei\ss ner and J.~Smejkal,
  Nucl.\ Phys.\ A {\bf 954}, 17 (2016).

\bibitem{Kamiya:2016} 
  Y.~Kamiya, K.~Miyahara, S.~Ohnishi, Y.~Ikeda, T.~Hyodo, E.~Oset and W.~Weise,
  Nucl.\ Phys.\ A {\bf 954}, 41 (2016).

\bibitem{Sada:2016} 
  Y.~Sada {\it et al.} [J-PARC E15 Collaboration],
  PTEP {\bf 2016}, 051D01 (2016). 

\bibitem{Sekihara:2016} 
  T.~Sekihara, E.~Oset and A.~Ramos,
  PTEP {\bf 2016}, 123D03 (2016).

\bibitem{Gal:2016}  
  A.~Gal, E.~V.~Hungerford and D.~J.~Millener,
  Rev.\ Mod.\ Phys.\  {\bf 88}, 035004 (2016). 
%
\bibitem{Shevchenko:2016} 
  N.~V.~Shevchenko,
  Few Body Syst.\  {\bf 58}, 6 (2017).

\bibitem{Bazzi:2011} 
  M.~Bazzi {\it et al.} [SIDDHARTA Collaboration],
  Phys.\ Lett.\ B {\bf 704}, 113 (2011). 

\bibitem{Oller:2000} 
  J.~A.~Oller and U.-G.~Mei\ss ner,
  Phys.\ Lett.\ B {\bf 500}, 263 (2001).

\bibitem{Oset:2002} 
  E.~Oset, A.~Ramos and C.~Bennhold,
  Phys.\ Lett.\ B {\bf 527}, 99 (2002)
  Erratum: [Phys.\ Lett.\ B {\bf 530}, 260 (2002)].

\bibitem{Ikeda:2012} 
  Y.~Ikeda, T.~Hyodo and W.~Weise,
  Nucl.\ Phys.\ A {\bf 881}, 98 (2012). 

\bibitem{Guo:2012}   
  Z.-H.~Guo and J.~A.~Oller,
  Phys.\ Rev.\ C {\bf 87}, 035202 (2013).

\bibitem{Mai:2012}   
  M.~Mai and U.-G.~Mei\ss ner,
  Nucl.\ Phys.\ A {\bf 900}, 51 (2013). 

\bibitem{Ohnishi:2013}   
  S.~Ohnishi, Y.~Ikeda, H.~Kamano and T.~Sato,
  Phys.\ Rev.\ C {\bf 88}, 025204 (2013). 

\bibitem{MG}
  A.~M\"uller-Groeling, K. Holinde, and J. Speth,
  Nucl. Phys. A {\bf 513}, 557 (1990).

\bibitem{Hai:2011}
  J.~Haidenbauer, G.~Krein, U.-G.~Mei{\ss}ner and L.~Tolos,
  Eur.\ Phys.\ J.\ A {\bf 47}, 18 (2011)

\bibitem{Kamano-Nakamura-Lee-Sato}
   H.~Kamano S.X.~Nakamura, T.-S.H.~Lee, and T.~Sato,
   Phys.\ Rev.\  C {\bf 90}, 065204 (2016).

\bibitem{Zmeskal:2015} 
  J.~Zmeskal {\it et al.},
  Acta Phys.\ Polon.\ B {\bf 46}, 101 (2015).
%
\bibitem{Moriya:2013} 
  K.~Moriya {\it et al.} [CLAS Collaboration], 
  Phys.\ Rev.\ C {\bf 87}, 035206 (2013). 
\bibitem{Moriya:2013A} 
  K.~Moriya {\it et al.} [CLAS Collaboration],
  Phys.\ Rev.\ C {\bf 88}, 045201 (2013). 
\bibitem{Lu:2013} 
  H.~Y.~Lu {\it et al.} [CLAS Collaboration], 
  Phys.\ Rev.\ C {\bf 88}, 045202 (2013).
\bibitem{Zychor:2007}   
  I.~Zychor {\it et al.},
  Phys.\ Lett.\ B {\bf 660}, 167 (2008). 
\bibitem{Agakishiev:2012} 
  G.~Agakishiev {\it et al.} [HADES Collaboration],
  Phys.\ Rev.\ C {\bf 87}, 025201 (2013).
\bibitem{Prakhov:2004} 
  S.~Prakhov {\it et al.} [Crystall Ball Collaboration],
  Phys.\ Rev.\ C {\bf 70}, 034605 (2004).

\bibitem{JparcE31}
   S.~Kawasaki {\it et al.},
   JPS Conf. Proc. {\bf 13}, 020018 (2017). 

\bibitem{JparcE31A}
   K.~Inoue {\it et al.}, presentation at the International workshop on Hadron and Nuclear Physics,
   Osaka, Japan, 12-14 March 2017, {\tt https://indico2.riken.jp/indico/\hfill\break
   contributionDisplay.py?contribId=43\&confId=2389}

\bibitem{Jparc} S. Ajimura et al., {\tt http://j-parc.jp/researcher/ \hfill\break 
   Hadron/en/pac\_1207/pdf/E31\_2012-9.pdf}

\bibitem{Revai:2014} 
  J.~Revai,
  Phys.\ Atom.\ Nucl.\  {\bf 77}, 509 (2014).

\bibitem{HYP2015}
   K.~Miyagawa and J.~Haidenbauer,
   JPS Conf. Proc. {\bf 17}, 072005 (2017). 

\bibitem{Ohnishi:2016} 
  S.~Ohnishi, Y.~Ikeda, T.~Hyodo and W.~Weise,
  Phys.\ Rev.\ C {\bf 93}, 025207 (2016). 

\bibitem{KM:2012}
   K.~Miyagawa and J.~Haidenbauer,
   Phys. Rev. {\bf 85}, 065201 (2012). 

\bibitem{Sekihara:2012} 
  J.~Yamagata-Sekihara, T.~Sekihara and D.~Jido,
  PTEP {\bf 2013}, 043D02 (2013).

\bibitem{Jido:2012} 
  D.~Jido, E.~Oset and T.~Sekihara,
  Eur.\ Phys.\ J.\ A {\bf 49}, 95 (2013).

\bibitem{Kamano-Lee}
   H.~Kamano and T.-S.H.~Lee,
   Phys.\ Rev.\  C {\bf 94}, 065205 (2016).

\bibitem{Cieply:2012} 
  A.~Ciepl\'y and J.~Smejkal, 
  Nucl. Phys. A {\bf 881}, 115 (2012).


\bibitem{generalP}
    W.~Gl\"ockle and K.~Miyagawa,
   Few-Body Syst. {\bf 30}, 241 (2001). 

\bibitem{Gloeckle96}
    W.~Gl\"ockle, H.~Witala, D.~H\"uber, H.~Kamada and J.~Golak,
    Phys. Rep. {\bf 274}, 107 (1996). 

\bibitem{Fong}
   R. Fong and J. Sucher,
   J. Math. Phys. {\bf 5}, 456 (1964).

\bibitem{relativisticND}
   H.~Witala, J.~Golak, W.~Gl\"ockle and H.~Kamada, 
   Phys.\ Rev.\  C {\bf 71}, 054001 (2005).

\bibitem{Newform}
   H.~Witala and W.~Gl\"ockle, 
   Eur.\ Phys.\ J.\ A {\bf 37}, 87 (2008).

\bibitem{Polyzou}
   B.~D.~Keister, W.~N.~Polyzou, 
   Phys.\ Rev.\  C {\bf 73}, 014005 (2006).

\bibitem{Kamada}
   H.~Kamada,  W.~N.~Polyzou,  H.~Witala and K.~Miyagawa,
   Few-Body Syst. {\bf 55}, 1079 (2014). 


\bibitem{Kaiser:1995} 
  N.~Kaiser, P.~B.~Siegel and W.~Weise,
  Nucl.\ Phys.\ A {\bf 594}, 325 (1995).

\bibitem{Cieply:2010} 
  A.~Ciepl\'y and J.~Smejkal,
  Eur.\ Phys.\ J.\ A {\bf 43}, 191 (2010).

\bibitem{KSU}
   H.~Zhang, J.~Tulpan,  M.~Shrestha, and D.M.~Manley, 
   Phys.\ Rev.\  C {\bf 88}, 035204 (2013).

\bibitem{Nij93NN} 
  V.~G.~J.~Stoks, R.~A.~M.~Klomp, C.~P.~F.~Terheggen and J.~J.~de Swart,
  Phys.\ Rev.\ C {\bf 49}, 2950 (1994).
 
\bibitem{Inoue:2017}        
  K.~Inoue {\it et al.},
  JPS Conf.\ Proc.\  {\bf 17}, 072003 (2017).
%
\bibitem{PDG} 
 C. Patrignani et al. [Particle Data Group], Chin. Phys. {\bf C} 40, 100001 (2016). 

\bibitem{Roca:2013} 
  L.~Roca and E.~Oset,
  Phys.\ Rev.\ C {\bf 88}, 055206 (2013).

\bibitem{Nakamura:2013} 
  S.~X.~Nakamura and D.~Jido,
  PTEP {\bf 2014}, 023D01 (2014). 

\bibitem{Mai:2014} 
  M.~Mai and U.-G.~Mei\ss ner,
  Eur.\ Phys.\ J.\ A {\bf 51}, 30 (2015). 

\bibitem{BB1}
   R.~Balian, E.~Br\'ezin,
   Nuovo Cimento  B {\bf 61}, 403 (1969).
\end{thebibliography}
\end{document}